\documentclass[10pt,a4paper]{article}
\usepackage{jheppub}
\usepackage{pdflscape}
\usepackage{amsmath}
\usepackage{amssymb}
\usepackage{dcolumn}
\usepackage{bm}

\usepackage{color}
\usepackage{epsfig}
\usepackage{amsfonts}
\usepackage{graphicx}
\usepackage{subfigure}
\usepackage{dcolumn}

\usepackage[dvipsnames]{xcolor}

\begin{document}
\title{ Quantum Work and Information Geometry of a Quantum Myers-Perry Black Hole}
\author[a]{Behnam Pourhassan,}
\author[b]{Salman Sajad Wani,}
\author[c]{Saheb Soroushfar,}
\author[d]{Mir Faizal}

\affiliation[a] {Canadian Quantum Research Center, 204-3002 32 Ave Vernon, BC V1T 2L7 Canada.}
\affiliation[a] {School of Physics, Damghan University, Damghan, 3671641167, Iran.}
\affiliation[b] {Canadian Quantum Research Center, 204-3002 32 Ave Vernon, BC V1T 2L7 Canada.}
\affiliation[c] {Faculty of Technology and Mining, Yasouj University, Choram 75761-59836, Iran.}
\affiliation[d] {Department of Physics and Astronomy, University of Lethbridge, Lethbridge, Alberta, T1K 3M4, Canada.}
\affiliation[d] {Irving K. Barber School of Arts and Sciences, University of British Columbia, Kelowna, British Columbia, V1V 1V7, Canada.}
\affiliation[d] {Canadian Quantum Research Center, 204-3002 32 Ave Vernon, BC V1T 2L7 Canada.}

\emailAdd{b.pourhassan@du.ac.ir}
\emailAdd{salmansajadwani@gmail.com}
\emailAdd{soroush@yu.ac.ir}
\emailAdd{mirfaizalmir@googlemail.com}

\abstract{In this paper, we will obtain quantum work for a quantum scale  five dimensional   Myers-Perry black hole.   Unlike  heat represented by Hawking radiation, the  quantum work  is represented by  a unitary information preserving process, and  becomes important for black holes only at small quantum scales. It will be observed that at such short distances, the quantum work will be corrected by non-perturbative quantum gravitational corrections. We will use the Jarzynski equality  to obtain this quantum work modified by  non-perturbative quantum gravitational corrections. These non-perturbative corrections will also modify the stability of a quantum Myers-Perry black hole. We will define a quantum corrected information geometry by incorporating the  non-perturbative quantum  corrections in  the information geometry of a  Myers-Perry black hole. We will use several different    quantum corrected effective  information metrics to analyze the stability of a quantum Myers-Perry black hole.}


\maketitle

\section{Introduction}
It is known that  black holes are  maximum entropy objects, and this maximum entropy of black holes scales with the area of their horizon   \cite{1a, 2a}. So, the  maximum entropy of  a region   scales with its area rather than its volume. This scaling of maximum entropy with the area of a region has motivated   the development of the  holographic principle \cite{4a, 5a}.
It has been suggested that the relation between area and entropy can be modified due to quantum gravitational corrections \cite{6ab, 7ab, 7ba}. However,  as the corrected entropy is still a function of the area of the horizon, these  corrections  can also be investigated   using  the  holographic principle     \cite{6a, 7a}. In fact, as AdS/CFT correspondence is a concrete realization of the holographic principle, it has been used to obtain such     quantum corrections to the entropy of  black holes \cite{18, 18a, 18b, 18c, 18d}.  Such quantum correction have also been obtained using
extremal limit of  black holes   \cite{19, 19a}.
The quantum corrections to the entropy of a black hole can be obtained from a conformal field theory, using the  density of microstates   associated with conformal blocks   \cite{Ashtekar}. As the Cardy formula can be used to obtain the entropy of a conformal field theory, it has been used to calculate the    quantum corrections to the black hole entropy  \cite{Govindarajan}.  The quantum corrected  entropy of a black hole has also been obtained using the Rademacher expansion \cite{29}.
Thus, various different approaches have been used to obtain quantum corrected entropy of black holes.

The  geometry of space-time   emerges  from  thermodynamics in the Jacobson formalism \cite{gr12}. So, it is expected that quantum fluctuation of space-time can be obtained from  the thermal fluctuations in the Jacobson formalism  \cite{gr14}. As it  is possible to obtain the perturbative thermal fluctuations to the thermodynamics of various black holes  \cite{32, 32a, 32b, 32c, 32d}, it has been argued that  such  perturbative thermal fluctuations can be used   to  obtain perturbative quantum fluctuations \cite{gr12, gr14}. In fact, perturbative quantum  corrections to various  black holes have been constructed using the perturbative thermal fluctuations   \cite{40a, 40b, 40c, 40d}. As the temperature of black holes scales inversely with its size, the thermal fluctuations can be neglected for large black holes. Since the quantum fluctuations can be  obtained using thermal fluctuations   \cite{gr12, gr14}, we can also neglect the quantum fluctuations of large black holes.   However, these quantum corrections become important for smaller   black holes, and  they cannot be neglected after a certain scale. The small perturbative quantum corrections to various black holes can be obtained using  thermal perturbations near the equilibrium   \cite{40a, 40b, 40c, 40d}. These perturbative corrections are valid for black holes, where the thermal fluctuations are small enough for them to be expressed as small   perturbations near the equilibrium.  Thus, they are expressed in terms  of the original equilibrium entropy and temperature  \cite{32, 32a, 32b, 32c, 32d}. However, for black holes whose size is   comparable to Planck scale, the corrections to the entropy cannot be described as perturbative corrections near the equilibrium. At such a short distance, we need to consider the full non-perturbative quantum gravitational corrections to the thermodynamics of black holes. It is expected  that such  non-perturbative quantum gravitational corrections produce non-trivial modifications  to the entropy of black holes. In fact, it has been demonstrated that the corrected entropy of such quantum  black holes can be expressed as an  exponential function of the original entropy  \cite{2007.15401}. Such non-perturbative corrections to the entropy of a black hole have also been  obtained from  the Kloosterman sums \cite{Dabholkar}. To obtain such non-perturbative corrections, the near horizon behavior of  mass-less supergravity fields was investigated using the AdS/CFT correspondence \cite{ds12, ds14}.
Now, as these   non-perturbative corrections \cite{2007.15401, Dabholkar}  have been analyzed  using    string theoretical effects    \cite{ds12, ds14}, it is important to investigate such non-perturbative corrections to the extra dimensional geometries motivated by string theory.

As string theory is only consistent in extra dimensions, it is possible to motivate the study of extra dimensional black holes from it. So, using such string theoretical motivation, it is possible to study a  five dimensional  Myers-Perry black hole \cite{MP}. The  Myers-Perry black hole  rotates in two independent planes, and can be considered as a generalization of the Kerr black hole to extra dimensions.  The thermodynamics of such a five dimensional  Myers-Perry black hole has been investigated \cite{thermo, thermo1}. The effect of perturbative quantum corrections to the thermodynamics of the Myers-Perry black hole has also been studied \cite{thermo4}.
So,  we will  analyze the effects of non-perturbative  corrections to a five dimensional quantum Myers-Perry black hole. As these non-perturbative corrections cannot be analyzed as small fluctuations near the equilibrium, it is important to  use non-equilibrium quantum  thermodynamics to analyze them \cite{adba0, adba1, adba2, adba4}.
It is possible to define a quantum analog of classical  work in quantum thermodynamics.  It may be noted that as quantum work is associated with a quantum process, it cannot be obtained using a  single  measurement of a quantum observable. However, it possible to obtain quantum work using the quantum Crooks fluctuation theorem,  by making two measurements \cite{work1, work2}. As it is possible to obtain  Jarzynski equality  from quantum Crooks fluctuation theorem, the Jarzynski equality has been  used to relates quantum work  with the difference of equilibrium free energies \cite{eq12, eq14}. Now as we can obtain the equilibrium free energies for a black hole, it can be used to obtain information about the quantum  work done during its evaporation.  Thus, the  Jarzynski equality has been used to obtain quantum work for black holes \cite{rz12, rz14}. It may be noted that here quantum work is  obtained  from the original Bekenstein-Hawking entropy of the black holes. This original Bekenstein-Hawking entropy is obtained from semi-classical approximations by using quantum field theory in curved space-time \cite{1a, 2a}. Hence, for the original
 Bekenstein-Hawking entropy  quantum gravitational effects are neglected.  However, at such scales, quantum gravitational corrections  cannot be neglected, and we need to consider the non-perturbative  quantum gravitational corrections \cite{2007.15401, Dabholkar} to quantum work.
So, in this paper, we will analyze the non-perturbative quantum gravitational corrections to the quantum work of a Myers-Perry black hole during its evaporation. It may be noted that the information geometry of the Myers-Perry black hole has also been studied using  different  information metrics \cite{thermo5}. We will analyze the non-perturbative quantum gravitational corrections to the information geometry of the Myers-Perry black hole, by defining quantum corrected effective informational metrics.

\section{Non-Perturbative Corrections}
In this section, we will analyze the non-perturbative corrections to a  Myers-Perry black hole.  The Myers-Perry black hole is    a five dimensional black hole, which rotates in two independent planes.
The metric for such  a five dimensional  Myers-Perry black hole can be expressed as  (in Planck units) \cite{MP, thermo},
\begin{eqnarray}\label{metric}
ds^{2}&=&-dt^{2}+\frac{\Sigma r^{2}}{\Delta}dr^{2}+\Sigma d\theta^{2}+(r^{2}+a^{2})\sin^{2}\theta d\varphi^{2}\nonumber\\
&+&r^{2}\cos^{2}\theta d\zeta^{2}+\frac{\mu}{\Sigma}\left(dt - a\sin^{2}\theta d\varphi\right)^{2},
\end{eqnarray}
where $a$ is angular momentum and $\mu$ is mass parameter. It may be noted that usually the general form of Myers-Perry black hole solution is parameterized by $m, a_1, a_2$, and here we have taken $m = \mu, a_1 =a, a_2 =0$, for simplification.   Here
$\Sigma= r^{2}+a^{2}\cos^{2}\theta, $ and  $\Delta= r^{2}\left(r^{2}+a^{2}-\mu\right)$, are suitable functions in the metric \cite{thermo}.
It is possible to express the radius of the  horizon in terms of $\mu$ which is a  parameter related to mass, and $a$ which is a  parameter related to the  angular momentum,   as  $r_{h}=\sqrt{\mu-a^{2}} $.
The explicit  mass and angular momentum of the Myers-Perry black hole can expressed in   terms of  $\mu$ and   $a$ as $M={3\mu}/{8}$ and $J = {2aM}/{3}$. We have expressed the  Myers-Perry black hole  using $(M, J)$ as the non-perturbative corrections to the information  geometry can be directly obtained using $(M, J)$.\\
The surface gravity of these solutions given by \cite{temp},
\begin{equation}\label{surface}
\kappa=\frac{r_{h}}{r_{h}^{2}+a^{2}}.
\end{equation}
The temperature of the Myers-Perry black hole can now be written as $T={\kappa}/{2\pi}$ \cite{temp}. By using the horizon radius, the Myers-Perry black hole temperature can obtain in terms of mass and angular momentum as,
\begin{equation}\label{temperature}
T=\frac{\sqrt{96M^{3}-81J^{2}}}{32\pi^{2}M^{2}}.
\end{equation}
In Fig. \ref{T} we can see behavior of temperature in terms of $M$ by variation of the angular momentum. We can see that there is a lower limit for $M$ in presence of $J$.

\begin{figure}[h!]
\begin{center}$
\begin{array}{cccc}
\includegraphics[width=70 mm]{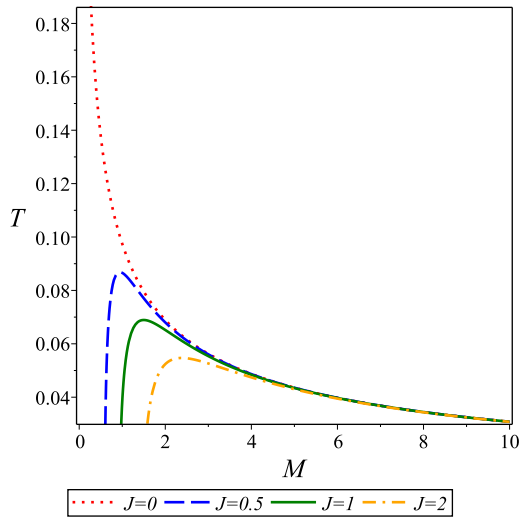}
\end{array}$
\end{center}
\caption{Temperature of the Myers-Perry black hole in terms of mass.}
\label{T}
\end{figure}

The rotation parameter has an  upper bound, which is given by  $a<\sqrt{{8M}/{3}}$. Now for quantum sized black holes,  we can write $M<1$. So,  using $J^2<\frac{96}{81}M^3$,   for  quantum sized black hole (with  $M<1$), we observe  $J\ll1$.
It may be noted that the entropy of a back hole can  be obtained from   microstates   of a conformal field theory  \cite{mi12, mi14,mi16, mi18}. The microstates  of black holes with angular momentum have been studied, and it has been observed  that they can be  described as a function of its mass and angular momentum $\Omega (M, J)$ \cite{ro12, ro14, ro16, ro18}. So,  we can write the entropy of a
Myers-Perry black hole  in terms of its microstates  as
\begin{equation}\label{SO}
S_{0}=\ln{\Omega} (M, J) = \frac{2\pi^{2}\sqrt{96M^{3}-81J^{2}}}{9}.
\end{equation}

This original  entropy of the  Myers-Perry black hole will get modified by quantum gravitational corrections at short distances. This is because the relation between entropy and area of the horizon of any  black hole (including a Myers-Perry black hole) is expected to be modified due to the quantum gravitational corrections at short distances.
It may be noted that in the Jacobson formalism,    geometry of space-time   emerges  from  thermodynamics  \cite{gr12}, and so these quantum corrections can be obtained from the thermal fluctuations   \cite{gr14}.
In fact, perturbative thermal fluctuations have been used to obtain  perturbative quantum corrections to the geometry of  black holes  \cite{40a, 40b, 40c, 40d}. Such corrections are universal, and can be obtained from   microstates  of a conformal field theory  \cite{mi12, mi14,mi16, mi18}. It has been demonstrated that perturbative correction to the entropy of a black hole can be obtained from   modular invariance. of the partition functions for a  conformal field theory \cite{mi12, mi14,mi16, mi18}, and scales as
\cite{32, 32a, 32b, 32c, 32d}
\begin{equation}
S_{per} =  f(\Omega)  \sim   \ln (S_0) + ....,
\end{equation}
where $S_0$ is the original equilibrium entropy of the black hole. These corrections to the original equilibrium entropy have been obtained by analyzing small fluctuations around the equilibrium.  This is done by first noting that from  modular invariance of the  partition function for the conformal field theory \cite{mi12, mi14,mi16, mi18}, we can write    $S(\beta) = a \beta^n + b \beta^m,$ with the constant  $a, b , n, m, >0$. Now as $S(\beta)$ has an extremum   $\beta_0 = (nb/ma)^{1/m+n} = T^{-1}$, these perturbations can be expressed in terms of the original equilibrium entropy $S_0 = S(\beta_0)$  (with  the original  equilibrium temperature $T = 1/\beta_0$) \cite{32, 32a, 32b, 32c, 32d}. These perturbative corrections become important when the black hole is small, and can be neglected when it is very large. However, as the black hole reduces further, this perturbative description does not   hold, and we need to consider non-equilibrium  corrections to the thermodynamics of black holes.
These non-equilibrium corrections will correspond to non-perturbative quantum corrections to the geometry of black holes. Such non-perturbative corrections would modify the original entropy of a black hole. In fact, it has been argued that  the non-perturbative corrections to the black hole would scale as  \cite{2007.15401, Dabholkar, ds12, ds14}
\begin{equation}\label{Sm}
S_{m}=\frac{1}{\Omega}\sim e^{-S_{0}}.
\end{equation}
So,  we will analyze the effects of such non-perturbative corrections  \cite{2007.15401, Dabholkar, ds12, ds14}  on a Myers-Perry black hole. We can express the corrected  entropy of a Myers-Perry black hole as $ S=S_{0}+ \eta e^{-S_{0}}$, by introducing a parameter $\eta$.  This parameter  controls the strength of the non-perturbative corrections, and has been  motivated from the use of such a control parameter in   perturbative corrections \cite{32, 32a, 32b, 32c, 32d}. The value of $\eta$ is fixed in such a way that for a large black hole, these corrections are negligible, and they only change the behavior of thermodynamics for   small quantum scale Myers-Perry black holes.  So, we can explicitly  write the   non-perturbatively  corrected entropy for such a small quantum scale Myers-Perry black holes as
\begin{equation}\label{SMJ}
S=\frac{2\pi^{2}\sqrt{96M^{3}-81J^{2}}}{9}+\eta \, \exp \left ({-\frac{2\pi^{2}\sqrt{96M^{3}-81J^{2}}}{9}}\right).
\end{equation}
This  exponential non-perturbative quantum correction to the original entropy of a  Myers-Perry black hole is neglected at large distances.  It has been argued that perturbative  correction to the entropy can produce interesting  modifications for  the thermodynamic behavior of the system \cite{32, 32a, 32b, 32c, 32d}. Thus, we expect that the non-perturbative correction \cite{2007.15401, Dabholkar, ds12, ds14} would also produce non-trivial modifications to the  quantum  thermodynamics of this system. So, we will be using  the non-equilibrium quantum thermodynamics \cite{adba0, adba1, adba2, adba4} to analyze such a system, as at such small quantum scales, with non-perturbative quantum corrections, non-equilibrium description has to be used to study a quantum  Myers-Perry black hole.

\section{Quantum Work}
 In previous section, we have analyzed the quantum gravitational corrections to a quantum sized Myers-Perry black hole. We also observed that at such  a scale, we cannot express the system as a equilibrium system, and we need to use  non-equilibrium quantum thermodynamics to investigate it.
 It is known that the quantum  work  is one of the  most important quantities  in quantum thermodynamics \cite{adba0, adba1, adba2, adba4}.
 The quantum work can be obtained from the quantum
Crooks fluctuation theorem using  Jarzynski  equality  \cite{work1, work2}. The  Jarzynski  equality  relates the non-equilibrium measurements of the work done on a system to difference between  equilibrium free energies. As it is possible to calculate the free energies between two thermodynamical states of a black hole,  the quantum work  done during the evaporation of a black holes has also been obtained  using the  Jarzynski  equality  \cite{rz12, rz14}. Even though the non-equilibrium  thermodynamics of black holes has been previously studied \cite{rz12, rz14}, the quantum gravitational corrections \cite{2007.15401, Dabholkar} to the non-equilibrium  thermodynamics were neglected  in those investigations.  However, at such small scales, at which the system cannot be described as an equilibrium system, we cannot neglect    quantum gravitational corrections to it \cite{2007.15401, Dabholkar}. 
Hence, we investigate the effect of such   quantum gravitational corrections \cite{2007.15401, Dabholkar} on quantum work  done during the evaporation of the Myers-Perry black hole. 

As the  Myers-Perry black hole evaporates, the number of  its  microstates   changes. So, a Myers-Perry black hole  with    initial  microstates  $\Omega(M_{2}, J_{2})$ can  evaporate to a  Myers-Perry black hole  with microstates  $\Omega(M_{1}, J_{1})$. This change is produced due to Hawking radiation, whose temperature scales with surface gravity as $T={\kappa}/{2\pi}$ \cite{temp}. Now when the black hole is large, its temperature is small, and the rate of change of  microstates is much smaller than the original microstates. However, as the black hole becomes small, its temperature increases, and the rate of change of mircostates also increases. Furthermore, as the microstates are proportional to the entropy, the total number of microstates also decreases as the black hole becomes small. Thus, at a sufficient small size, we cannot neglect the effect of the rate of change of microstates, and at this stage, we cannot consider the system in equilibrium. It may be noted that for Myers-Perry black hole, large is defined in a thermodynamic sense, as the microstates are functions of both mass and angular momentum, and their combined effect defines the size of thermodynamic system.  So, for small quantum sized black holes, we have to explicitly analyze the effects of such changes between two thermodynamic states, and use non-equilibrium quantum thermodynamics to analyze it  \cite{rz12, rz14}. 

As the microstates  are related to the entropy, at quantum scales,  we need to investigate the change in the  entropy of the Myers-Perry black hole due to a change in its microstates.  
Now at small scales, we have to also consider the  effects of non-perturbative quantum gravitational corrections to the entropy of such a  Myers-Perry  black hole  \cite{2007.15401, Dabholkar, ds12, ds14}. So, we can express the  change in the  quantum gravitationally corrected entropy of a Myers-Perry  black hole as $\Delta S=S(M_{2}, J_{2})-S(M_{1}, J_{1})$. This  can be explicitly written as
\begin{eqnarray}
\Delta S&=&\frac{2\pi^{2}}{9}\left[\sqrt{96M_{2}^{3}-81J_{2}^{2}}-\sqrt{96M_{1}^{3}-81J_{1}^{2}}\right]\nonumber \\ && +\eta \left[\exp \left({-\frac{2\pi^{2}\sqrt{96M_{2}^{3}-81J_{2}^{2}}}{9}}\right)-\exp\left({-\frac{2\pi^{2}\sqrt{96M_{1}^{3}-81J_{1}^{2}}}{9}}\right)
\right].
\end{eqnarray}
This change in the entropy can be used to obtain a change in the internal energy of this  Myers-Perry black hole.   This can be done by first noting  that the internal energy  of quantum Myers-Perry black holes (where we have neglected $\mathcal{O}(J^{4})$ terms), can be expressed as
\begin{eqnarray}\label{U-sol}
U&\approx&\frac{-128M^{4}+297MJ^{2}}{-128M^{3}+378J^2}\nonumber\\
&-&\frac{\eta}{28}\frac{-1792\pi^{2}M^{6}+4158\pi^{2}J^{2}M^{3}+243\sqrt{6}J^{2}M^{\frac{3}{2}}}{\pi^{2}M^{2}(-64M^{3}+189J^{2})}.
\end{eqnarray}
Then, we can define the change in internal energy as $\Delta U=U(M_{2}, J_{2})-U(M_{1}, J_{1})$. So,  we can express such a change in the internal energy of a Myers-Perry black hole evaporates as
\begin{eqnarray}\label{delta-U-sol}
\Delta U&=&\frac{-128M_{2}^{4}+297M_{2}J_{2}^{2}}{-128M_{2}^{3}+378J_{2}^2}-\frac{-128M_{1}^{4}+297M_{1}J_{1}^{2}}{-128M_{1}^{3}+378J_{1}^2}\nonumber\\
&-&\frac{\eta}{28}\frac{-1792\pi^{2}M_{2}^{6}+4158\pi^{2}J_{2}^{2}M_{2}^{3}+243\sqrt{6}J_{2}^{2}M_{2}^{\frac{3}{2}}}{\pi^{2}M_{2}^{2}(-64M_{2}^{3}+189J_{2}^{2})}\nonumber\\
&+&\frac{\eta}{28}\frac{-1792\pi^{2}M_{1}^{6}+4158\pi^{2}J_{1}^{2}M_{1}^{3}+243\sqrt{6}J_{1}^{2}M_{1}^{\frac{3}{2}}}{\pi^{2}M_{1}^{2}(-64M_{1}^{3}+189J_{1}^{2})}.
\end{eqnarray}

This quantity does not vanish, and  the internal energy of the Myers-Perry black hole changes, as  it evaporates from $\Omega(M_{2}, J_{2})$ to $\Omega(M_{1}, J_{1})$. Now it  is important to analyze the amount of quantum  work done as the internal energy of the Myers-Perry black hole changes from $U(M_{2}, J_{2})$ to $U(M_{1}, J_{1})$. This is because apart from Hawking radiation,  a part of the internal energy of the black hole does quantum work during its evaporation. It may be noted that for equilibrium processes, where we can neglect  change in the area of the black hole, we can also neglect this quantum work term. However, for quantum sized black holes, the change in the area is of the same order  as the area, and we cannot neglect the quantum work done as the black hole evaporates. So, change in the  total energy of the black hole is given by the energy radiated by the Hawking radiation, and the energy spend in doing quantum work $W$, as the black hole contracts.  Now let the total amount of heat in  Hawking radiation be denoted by $Q$,  as the black hole evaporates from  $\Omega(M_{2}, J_{2})$ to $\Omega(M_{1}, J_{1})$, then we can write \cite{10th}
\begin{equation}
\Delta U =  Q - \langle W \rangle
\end{equation}
where $\langle W \rangle$ is the average quantum work done as the black hole evaporates from  $\Omega(M_{2}, J_{2})$ to $\Omega(M_{1}, J_{1})$.
We can use the Jarzynski equality   to obtain this quantum  work done as the Myers-Perry  black hole  evaporates from $\Omega(M_{2}, J_{2})$ to $\Omega(M_{1}, J_{1})$  \cite{work1, work2}. Thus, using the  Jarzynski equality \cite{eq12, eq14}, we can write the quantum work in terms of   difference of the equilibrium  free energies of these two states \cite{rz12, rz14}
\begin{equation}
\langle e^{-\beta W} \rangle = e^{\beta \Delta F}.
\end{equation}
So, it is possible to relate  the quantum work done by the Myers-Perry  black hole to the difference of free energies   $F(M_{2}, J_{2})$ and  $F(M_{1}, J_{1})$. It is important to note that unlike heat (represented by Hawking radiation), the  quantum work done on a system is represented  by a unitary information preserving process \cite{12th, 12tha}. Thus,  there is an unitary  information preserving process associated with the evaporation of black holes. This unitary  information preserving process, which corresponds to quantum work, can be neglected at large scales, but cannot be neglected at quantum scales. It could be possible that information can leak out of a black hole due to such a process. So,  it is possible that the black hole information loss paradox \cite{paradox1, paradox2} occur  only     due to the use of   equilibrium thermodynamics to describe the evaporation of a black hole down to Planck scale. However, at such a small scale, we have to analyze the system using   non-equilibrium  quantum thermodynamics. As  quantum work, which is represented by an unitary  information preserving process, becomes important at that scale, it is possible that black hole information paradox can be resolved by the use of  non-equilibrium quantum thermodynamics. It would be interesting to use the relation between quantum information theory and non-equilibrium quantum thermodynamics to investigate the black hole information paradox \cite{info1, info2}. This can be done by using the results  obtained  in this  paper, along with the relation between quantum work and information theory \cite{info1, info2}. 
The presence of this unitary  information preserving process   might be the reason that the black hole information paradox does  not occur in gauge gravity duality  \cite{paradox4, paradox5}. So, it would be interesting to explicitly analyze the holographic dual to quantum work, and its consequences on black hole information paradox. However, here we will only analyze the effect of quantum gravitational corrections on quantum work for a quantum sized  Myers-Perry black hole. 

Now, we can obtain the quantum work corrected by non-perturbative corrections using quantum corrected free energy.  We can write this modified free energy, which has been corrected by non-perturbative quantum gravitational corrections as
\begin{eqnarray}\label{FMP}
F&=&\frac{M}{3}+\frac{9}{16}\frac{J^{2}}{M^{2}}\nonumber\\
&+&\frac{\eta}{2\pi^{2}}\int{\left(\frac{\sqrt{96M^{3}-81J^{2}}}{8M^{3}}-\frac{9}{\sqrt{96M^{3}-81J^{2}}}\right)e^{-\frac{2\pi^{2}\sqrt{96M^{3}-81J^{2}}}{9}}dM}.
\end{eqnarray}
Now, for a quantum scale  Myers-Perry  black hole, we  can approximate (neglecting $\mathcal{O}(J^{4})$) the free energy as
\begin{eqnarray}\label{FMP-sol}
F&\approx&\frac{128M^{3}+135J^{2}}{384M^{2}}\nonumber\\
&-&\frac{\eta}{2688M^{\frac{7}{2}}\pi^{2}}\left[896M^{\frac{7}{2}}\pi^{2}+945M^{\frac{3}{2}}\pi^{2}J^{2}+336\sqrt{6}M^3-81\sqrt{6}J^{2}\right].
\end{eqnarray}
So, as the Myers-Perry  black hole with microstates   $\Omega(M_1, J_1)$   evaporates to  microstates  $\Omega(M_2, J_2)$,  the change of free energy can be expressed as   $\Delta F=F(M_{2}, J_{2})-F(M_{1}, J_{1})$. Thus,  using  Jarzynski equality,  we can write the quantum work as
\begin{eqnarray}\label{DeltaFMP-sol}
\langle e^{-\beta W} \rangle &=& e^{\beta  \Delta F}\nonumber \\ &=& \exp \beta\left[ \frac{128M_{2}^{3}+135J_{2}^{2}}{384M_{2}^{2}}-\frac{128M_{1}^{3}+135J_{1}^{2}}{384M_{1}^{2}}\right. \nonumber\\
&&-\frac{\eta}{2688M_{2}^{\frac{7}{2}}\pi^{2}}\left[896M_{2}^{\frac{7}{2}}\pi^{2}+945M_{2}^{\frac{3}{2}}\pi^{2}J_{2}^{2}
+336\sqrt{6}M_{2}^3-81\sqrt{6}J_{2}^{2}\right]\nonumber\\
&&\left. +\frac{\eta}{2688M_{1}^{\frac{7}{2}}\pi^{2}}
\left[896M_{1}^{\frac{7}{2}}\pi^{2}+945M_{1}^{\frac{3}{2}}\pi^{2}J_{1}^{2}+336\sqrt{6}M_{1}^3-81\sqrt{6}J_{1}^{2}\right]\right].
\end{eqnarray}
Thus, we have obtained quantum work for a Myers-Perry black hole.  We observe that it depends on the strength of non-perturbative corrections.

We can write the partition function for a Myers-Perry black using its  microstates   \cite{ro16}. So, as the microstates  change from $\Omega(M_1, J_1)$ to $\Omega(M_2, J_2)$ during evaporation,  the partition function   changes  from  $Z_1 [\Omega(M_1, J_1)]$ to $Z_2[\Omega(M_1, J_2)]$. We can relate these partition functions for a Myers-Perry  black hole to the  quantum work done as the black hole evaporates between these two microstates . So, if the black hole with a partition function  $Z_1$ evaporates to a black hole with a partition function   $Z_2$, then  we can use  the Jarzynski equality to obtain
\cite{eq12, eq14}
\begin{equation}
\langle e^{-\beta W} \rangle \nonumber=\frac{Z_2}{Z_1}
\end{equation}
As $Z_2/Z_1$ has been expressed in terms quantum work, we can also express the relative weights of the partition function in terms of the difference between the  equilibrium  free energies as $  \exp({\beta \Delta F}) ={Z_2}/{Z_1}$.

Now, we can use Jensen inequality to relate  the average of exponential of quantum  work to exponential  of the average of quantum work for a Myers-Perry  black hole as
$ e^{\langle -\beta {W} \rangle } \leq \langle e^{-\beta W} \rangle
$. So, by using this inequality we  can obtain an inequality for the  quantum  work done during the evaporation of a Myers-Perry black hole. This  can be written in terms of the difference of the free energy, which in term can be written in terms of the masses $M_1, M_2$ and angular momentum $J_1, J_2$ as
 \begin{eqnarray}
 \langle  W \rangle &\geq & \Delta F\nonumber\\
  &=& \frac{128M_{2}^{3}+135J_{2}^{2}}{384M_{2}^{2}}-\frac{128M_{1}^{3}+135J_{1}^{2}}{384M_{1}^{2}} \nonumber\\
&-&\frac{\eta}{2688M_{2}^{\frac{7}{2}}\pi^{2}}\left[896M_{2}^{\frac{7}{2}}\pi^{2}+945M_{2}^{\frac{3}{2}}\pi^{2}J_{2}^{2}
+336\sqrt{6}M_{2}^3-81\sqrt{6}J_{2}^{2}\right]\nonumber\\
&+&\frac{\eta}{2688M_{1}^{\frac{7}{2}}\pi^{2}}
\left[896M_{1}^{\frac{7}{2}}\pi^{2}+945M_{1}^{\frac{3}{2}}\pi^{2}J_{1}^{2}+336\sqrt{6}M_{1}^3-81\sqrt{6}J_{1}^{2}\right].
 \end{eqnarray}

\begin{figure}[h!]
\begin{center}$
\begin{array}{cccc}
\includegraphics[width=60 mm]{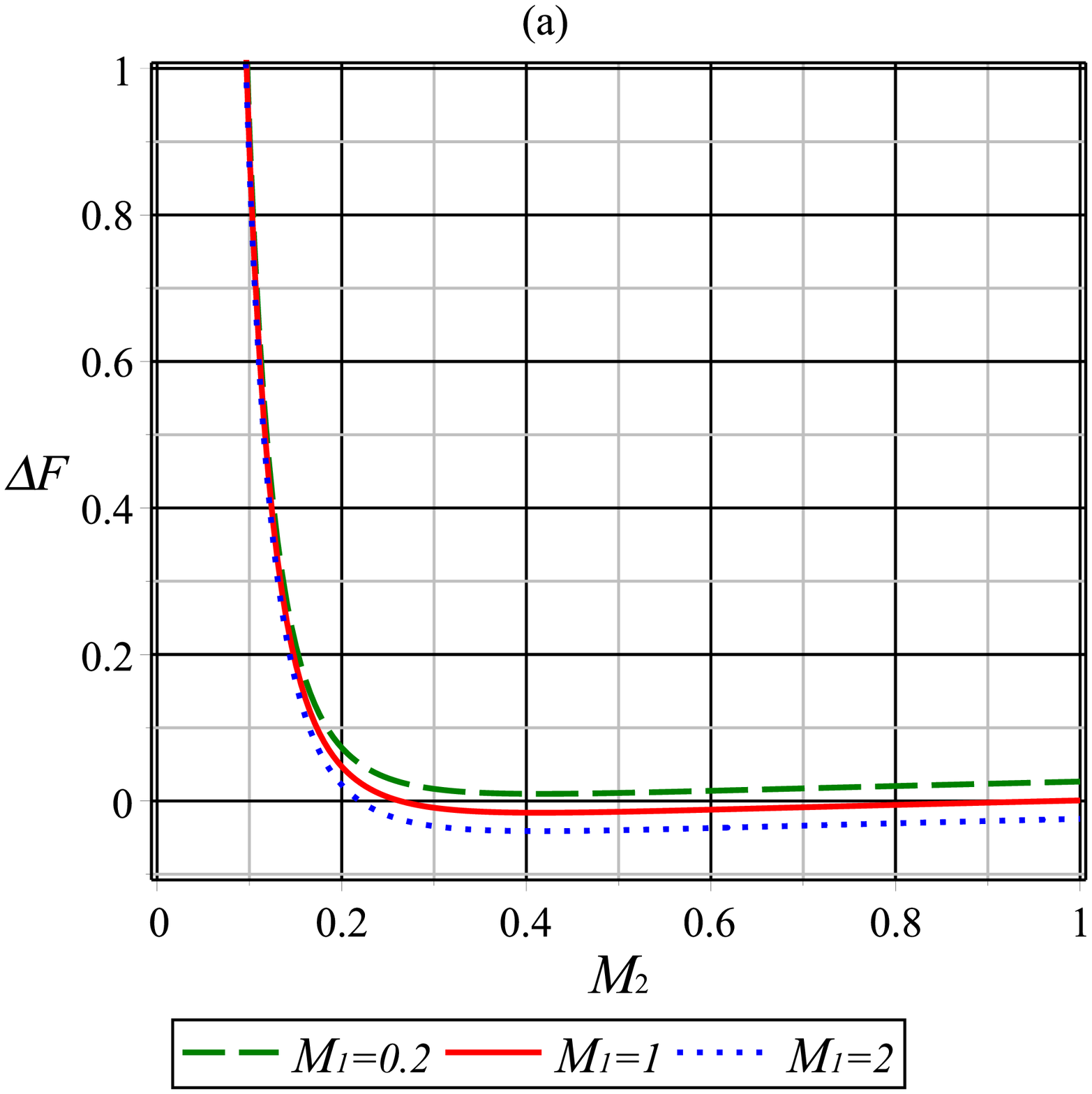}\includegraphics[width=60 mm]{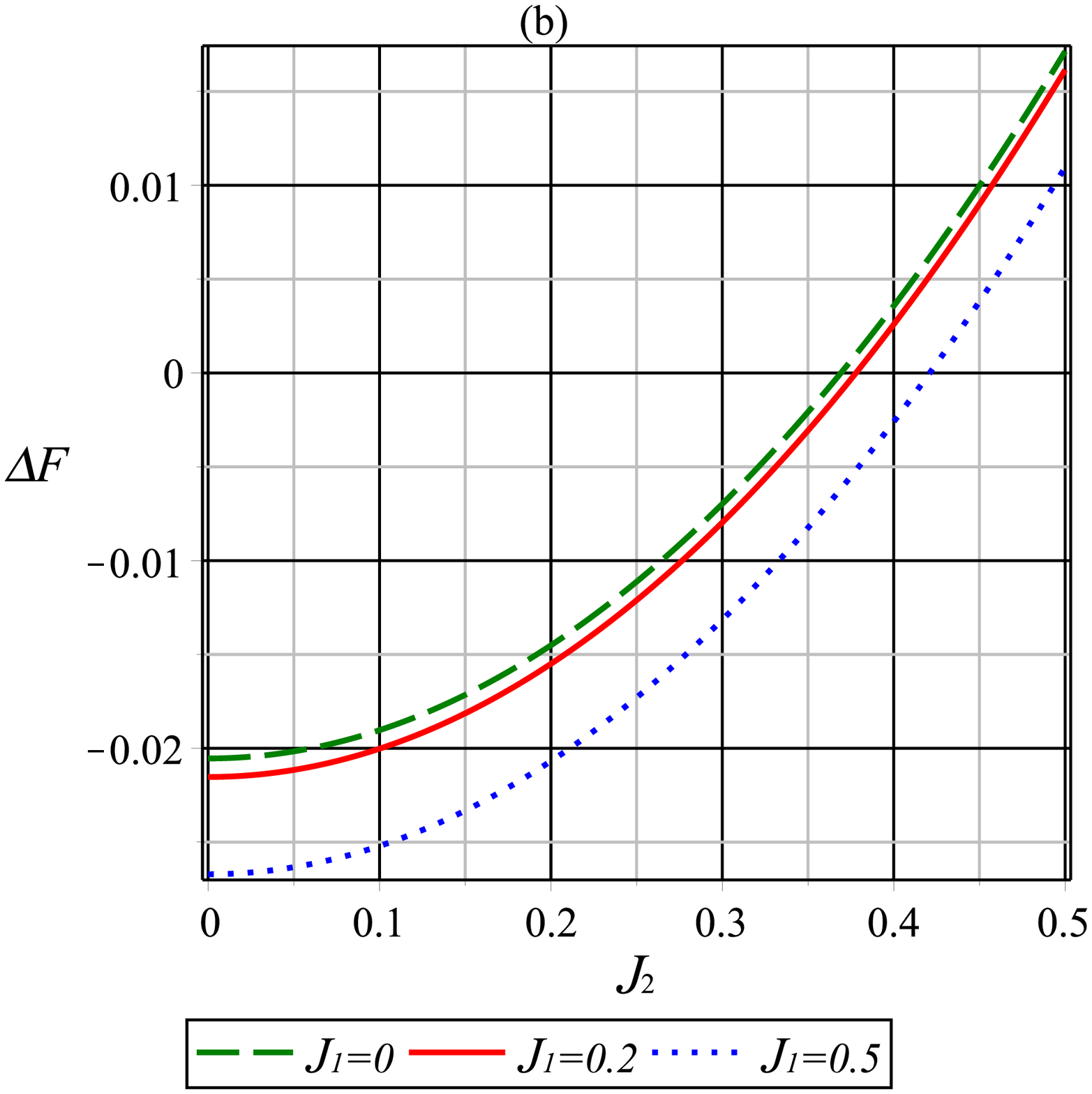}
\end{array}$
\end{center}
\caption{$\Delta F$ of the Myers-Perry black hole in terms of (a) final mass with $\eta=0.95$, $J_{1}=0.1$, and $J_{2}=0.2$; (b) final angular momentum  with $\eta=0.95$, $M_{1}=0.5$, and $M_{2}=1$.}
\label{figdeltaF2a}
\end{figure}

\begin{figure}[h!]
\begin{center}$
\begin{array}{cccc}
\includegraphics[width=60 mm]{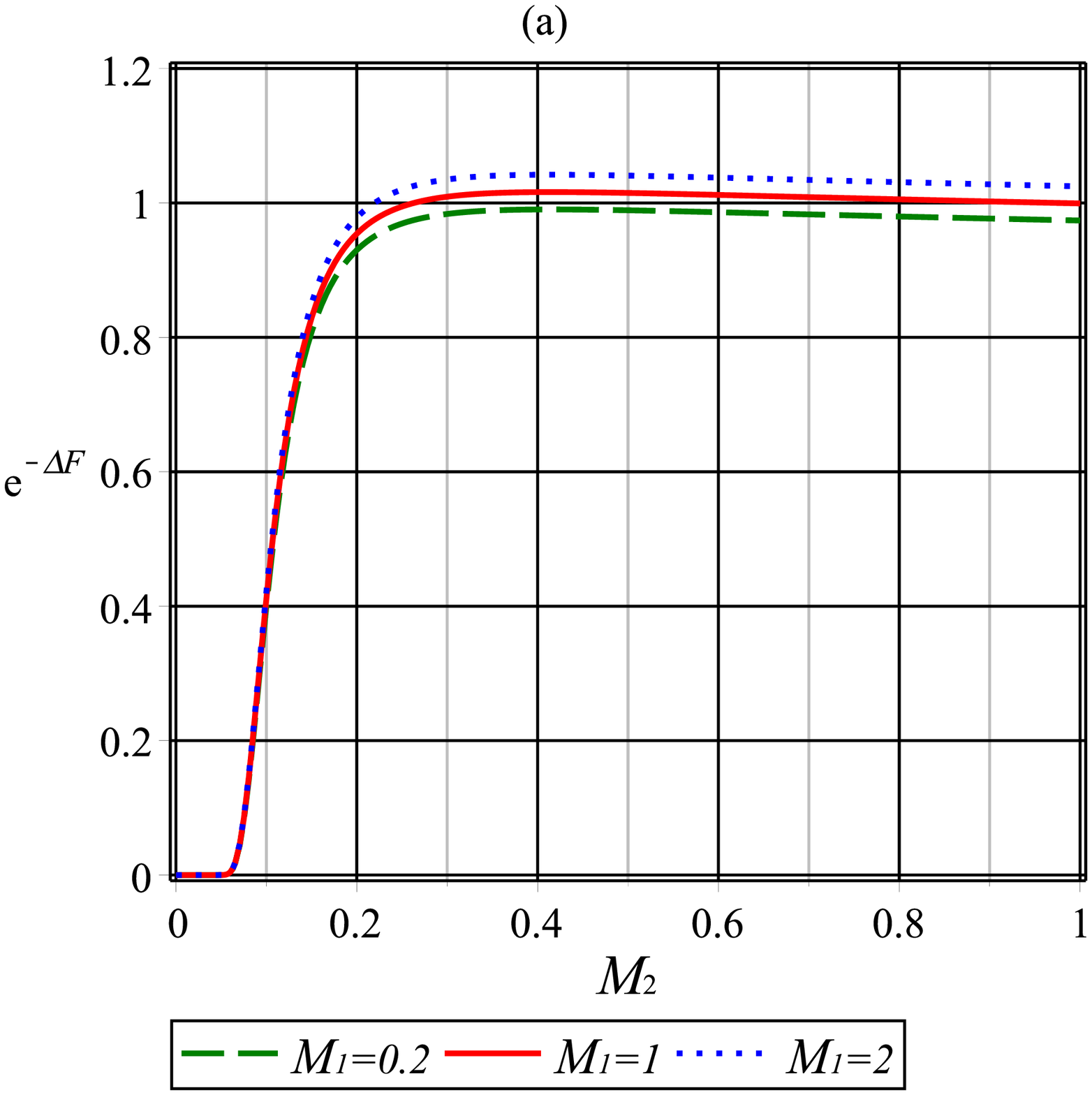}\includegraphics[width=60 mm]{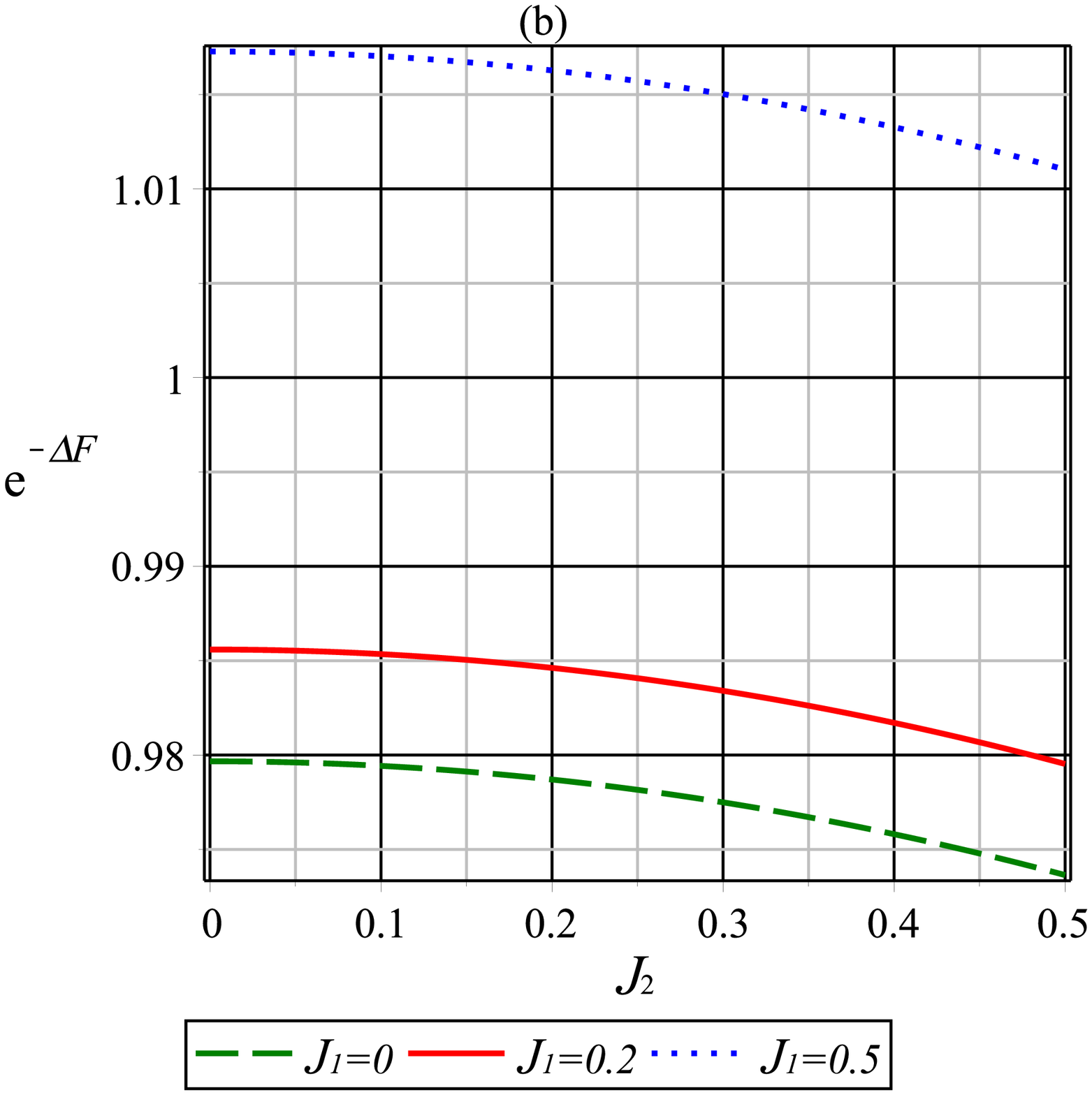}
\end{array}$
\end{center}
\caption{$e^{-\Delta F} =Z_2/Z_1 $ of the Myers-Perry black hole in terms of in terms of (a) final mass with $\eta=0.95$, $J_{1}=0.1$, and $J_{2}=0.2$; (b) final angular momentum  with $\eta=0.95$, $M_{1}=0.5$, and $M_{2}=1$.}
\label{figdeltaF2}
\end{figure}

Here, $\langle W\rangle = \Delta F$  for  processes at equilibrium.  It may be noted that in  Fig.~\ref{figdeltaF2a}, we have analyzed the  effect of non-perturbative corrections on the difference of free energy between two states  of a Myers-Perry black hole. This was done by first  analyzing the variation of the difference of free energy with mass, and then analyzing the variation of difference of  free energy with angular momentum.  As the  quantum work is bounded in terms of the free energy, we can view this as the correction to the bound on the    quantum work corrected by non-perturbative corrections.\\
In the Fig.~\ref{figdeltaF2}, we will investigate the effects of non-perturbative  corrections to the relative weights of the partition functions. As these relative weights can be expressed in terms of the difference of free energies, and this difference is corrected by non-perturbative corrections, these weights are also corrected by non-perturbative corrections. It may be noted that  weights can also be   related to the  exponential of quantum work using the Jarzynski equality \cite{eq12, eq14}. Here, again we investigate the variation of these relative weights with the difference of mass and difference of angular momentum of the Myers-Perry black hole. We can observe that non-perturbative quantum effects can produce interesting modifications to the relative weights of such partition functions.

\begin{figure}[h!]
\begin{center}$
\begin{array}{cccc}
\includegraphics[width=55 mm]{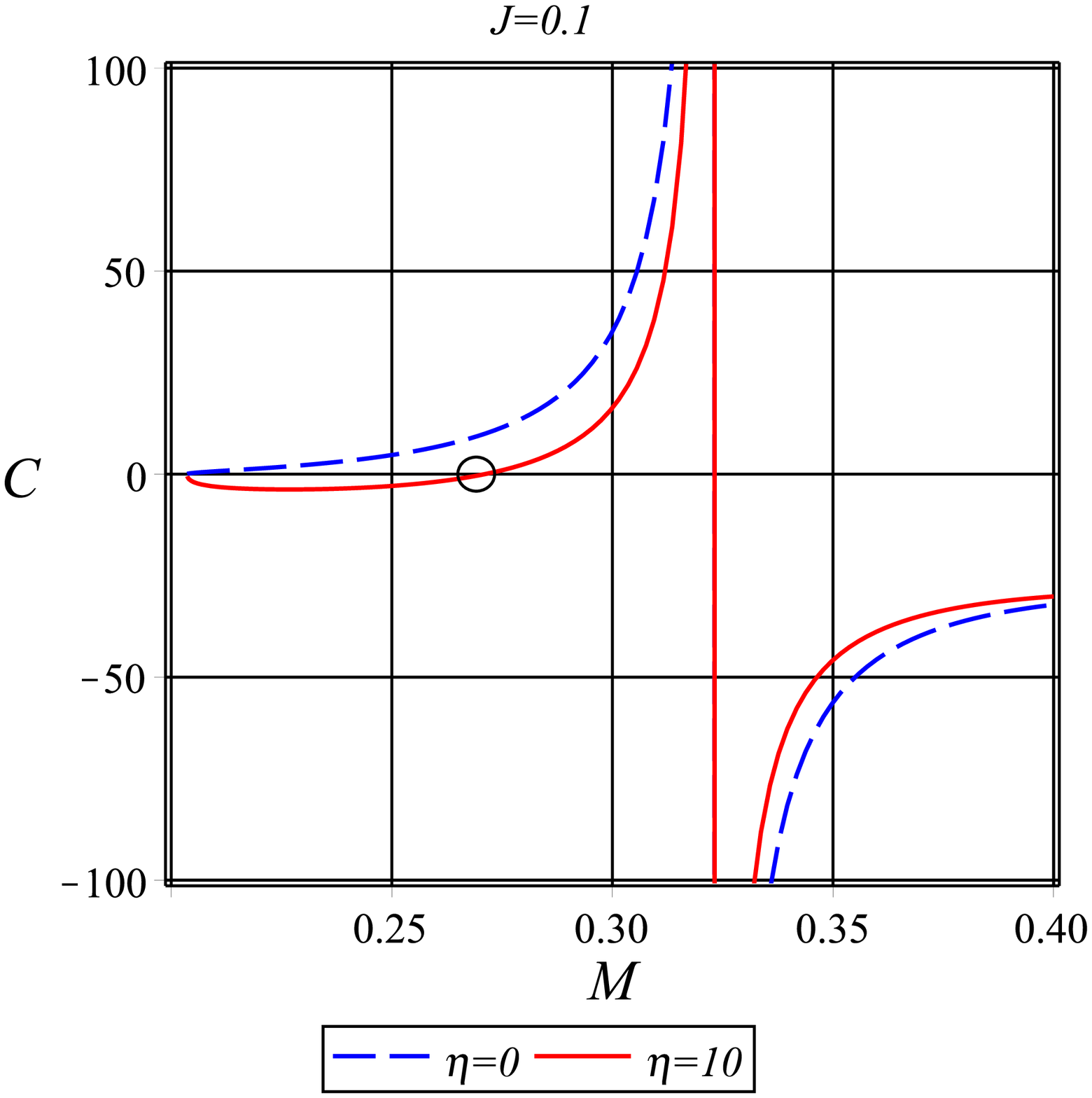}\includegraphics[width=55 mm]{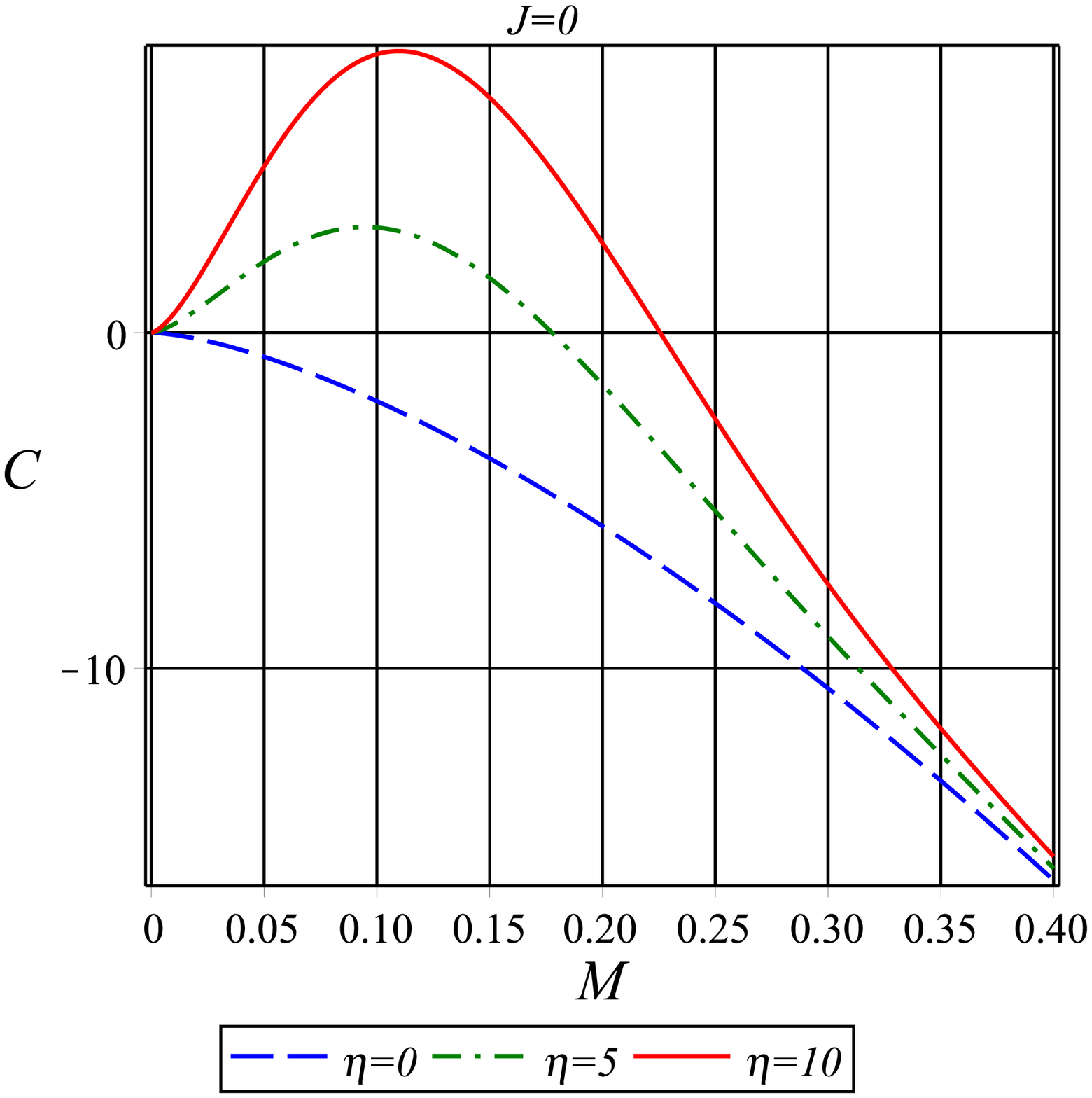}
\end{array}$
\end{center}
\caption{Specific heat of the Myers-Perry black hole in terms of mass.}
\label{fig1}
\end{figure}

As the quantum work is done during  the evaporation of a Myers-Perry black hole, it  is important to  analyze the effects of these non-perturbative corrections to the stability of these quantum scale Myers-Perry black hole. This can be done by analyzing the  non-perturbative corrections to  the specific heat of a Myers-Perry black hole
\begin{equation}\label{CMP}
C=\frac{16\pi^{2}M^{3}\sqrt{96M^{3}-81J^{2}}\left(1-\eta e^{-\frac{2\pi^{2}\sqrt{96M^{3}-81J^{2}}}{9}}\right)}{81J^{2}-24M^{3}}.
\end{equation}
The information about the  stability of this thermodynamics system can be obtained from the sign of this corrected specific heat. We plot the specific heat for the different values of $M$ in Fig. \ref{fig1}. We observe  that the original  Myers-Perry black hole with  large value of  $M$ is in an  unstable phase. However, there is a  minimum mass, at which   specific heat  vanish $C=0$. So, at this stage the black hole does not exchange energy with the surroundings. Furthermore, at this stage $T =0$,  and the  Myers-Perry black hole   stops radiating Hawking radiation. So, at this stage it forms  a stable  Myers-Perry black remnant. The non-perturbative quantum corrections change this behavior, as can  be observed from   the left part of plot in  Fig. \ref{fig1}. So, due to the non-perturbative quantum corrections,  the   specific heat of a   quantum scale  Myers-Perry black hole becomes negative, and it becomes  unstable.

At large scales, the original  Myers-Perry black hole in an unstable phase. At such a large scale, the   effects from  non-perturbative quantum corrections can be neglected. However, a first order phase transition occurs at a critical scale, for the original  Myers-Perry black hole. After this phase transition, the  original  Myers-Perry black is in  a stable phase.  The black hole mass at this critical point, where such a phase transition occurs is given by
\begin{equation}\label{criticalM}
M_{c}=\left[\frac{81}{96}\left(J^{2}+\left(\frac{\ln{\eta}}{2\pi^{2}}\right)^{2}\right)\right]^{1/3}.
\end{equation}

It is denoted by a circle line in left plot of Fig. \ref{fig1}. However, at such a small quantum  scales ($M\leq M_{c}$), non-perturbative quantum corrections  become important, and cannot be neglected. We observe that due to  these non-perturbative quantum corrections,  the  Myers-Perry black hole becomes  unstable, at such small scales.
It is interesting to note that the situation is different, when the angular momentum is neglected,  $J=0$. Here, the original black hole (Schwarzschild black hole) is initially  unstable, and  the  non-perturbative corrections make it   stable at small scales, as  illustrated by right plot of Fig. \ref{fig1}. Thus, we see that the behavior of  quantum corrected  specific heat depends on both    $J$ and $M$. We also observe that  the stability of the  Myers-Perry black holes is non-trivially modified by the  non-perturbative quantum corrections at small scales.

\section{Information  Geometry}\label{subr2}
It is possible to investigate  the stability of black holes using information geometry \cite{1111, 2222}.  The  phase transitions for black holes  can be investigated using  divergences of the Ricci scalar of the information geometry \cite{point1, point2, point4, point5, badpa, 3dmpl}. This Ricci scalar is obtained using from a specific form of the information metric. There are different information theoretical metrics, which represent different amount of information about a thermodynamic system.  In fact, it has been observed that the information about   phase transition of different black holes  can only be represented by certain  informational theoretical metrics. Thus, it is important to use various information theoretical metrics to analyze the phase transition in a black hole. The phase transition in a charged Gauss-Bonnet AdS black holes has been studied using a Ruppiner metric  \cite{12ra}, and the phase transition in a   Park black hole has been studied using a Weinhold metric \cite{12rw}.
However, it has been demonstrated that both the  Weinhold and Ruppiner metrics  cannot be used to obtain information about the  phase transition of a charged AdS black hole with a global monopole \cite{12qb}. It is possible to obtain such  information about the phase transition of such a  charged AdS black hole with a global monopole from    Quevedo and  HPEM metrics \cite{12qb}. The phase transition of   a black hole surrounded by the perfect fluid in Rastall theory has been studied using  the Ruppiner and HPEM metrics,  and it has been observed  such a phase transition  cannot be analyzed using   Weinhold and Quevedo metrics  \cite{rh12}. It has been demonstrated that for such a  black hole surrounded by the perfect fluid in the Rastall theory, it is possible to extract more information  from the HPEM metrics than the Ruppiner metric \cite{rh12}. Thus,  different amount of physical information about the phase transition  can be extracted from different information theoretical metrics.  

We have observed that at quantum scales, the quantum gravitational corrections cannot be neglected. These quantum gravitational corrections are expected to modify the information theoretical metrics for a black hole. So, we need to modify the original formalism of the information  geometry to incorporate these quantum gravitational effects.
As these information theoretical metrics are expressed in terms of the mass of a black hole, we need to  define a  novel quantum   mass of a Myers-Perry  black hole to construct  effective quantum corrected information metrics. This can be done by directly incorporating   the effect of non-perturbative quantum gravitational corrections into the quantum mass of a Myers-Perry  black hole. 
Furthermore, as the  quantum gravitational corrections can   be neglected for a Myers-Perry black hole at large scales,     this quantum    mass should  reduces to the original mass of the Myers-Perry black hole at such large scales. 

\begin{figure}[h]
	\centering
	\subfigure[$ J=0.01 $]{
		\includegraphics[width=0.50\textwidth]{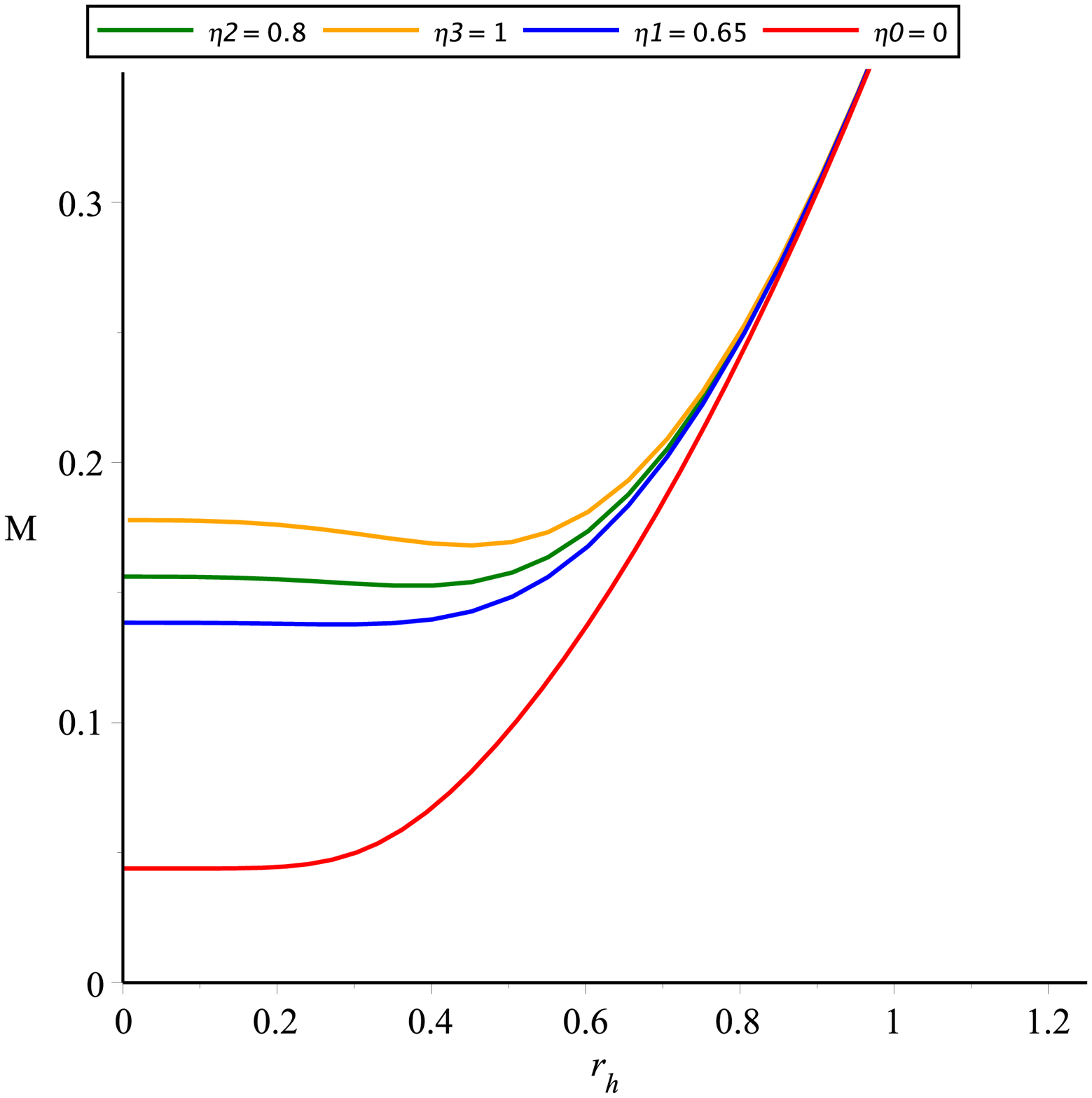}
	}
	\subfigure[$\eta=0.95$]{
		\includegraphics[width=0.50\textwidth]{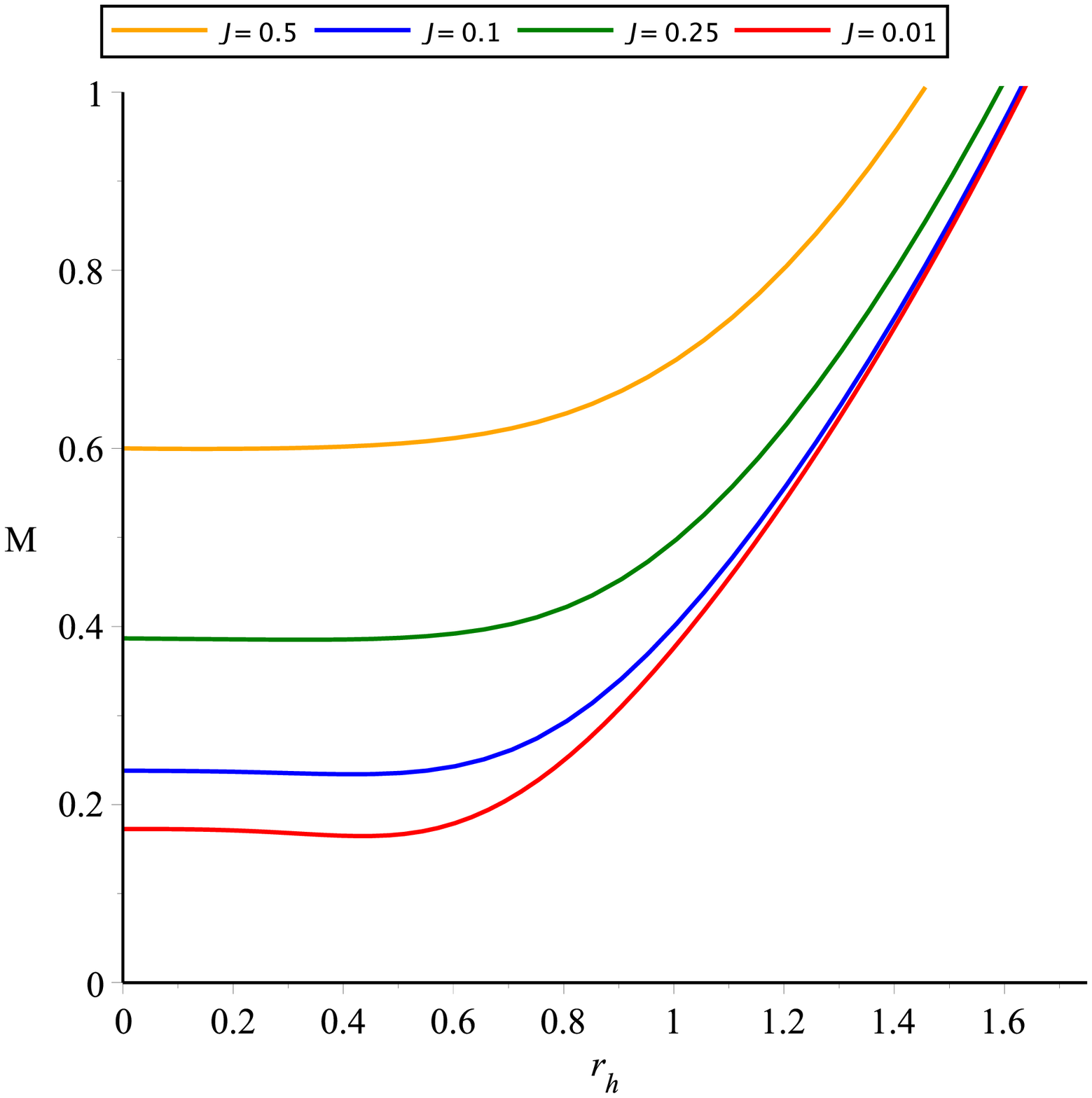}
	}
	\caption{Variations of mass in terms of horizon radius $ r_{h} $, for the Myers-Perry black hole in five dimensions.}
	\label{pic:M}
\end{figure}

\begin{figure}[h]
	\centering
	\subfigure[$ J=0.01 $]{
		\includegraphics[width=0.50\textwidth]{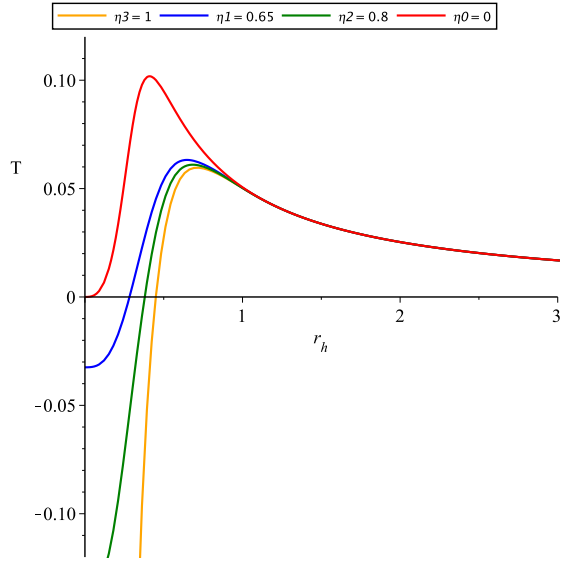}
	}
	\subfigure[$\eta=0.95$]{
		\includegraphics[width=0.50\textwidth]{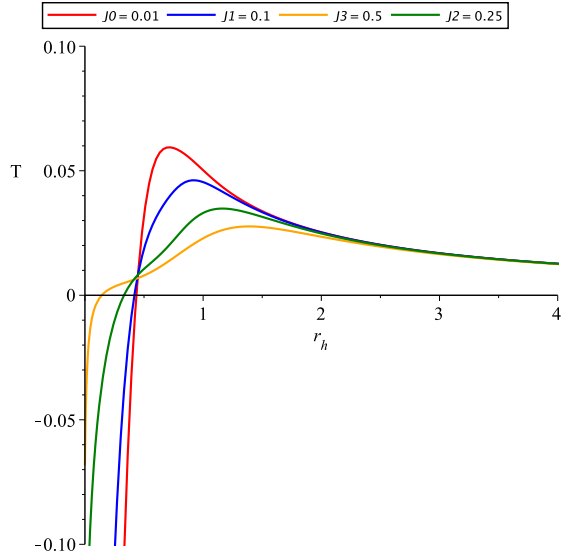}
	}
	\caption{Variations of temperature in terms of horizon radius $ r_{h} $, for the Myers-Perry black hole in five dimensions.}
	\label{pic:T}
\end{figure}

\begin{figure}[h]
	\centering
	\subfigure[$ J=0.01 $]{
		\includegraphics[width=0.4\textwidth]{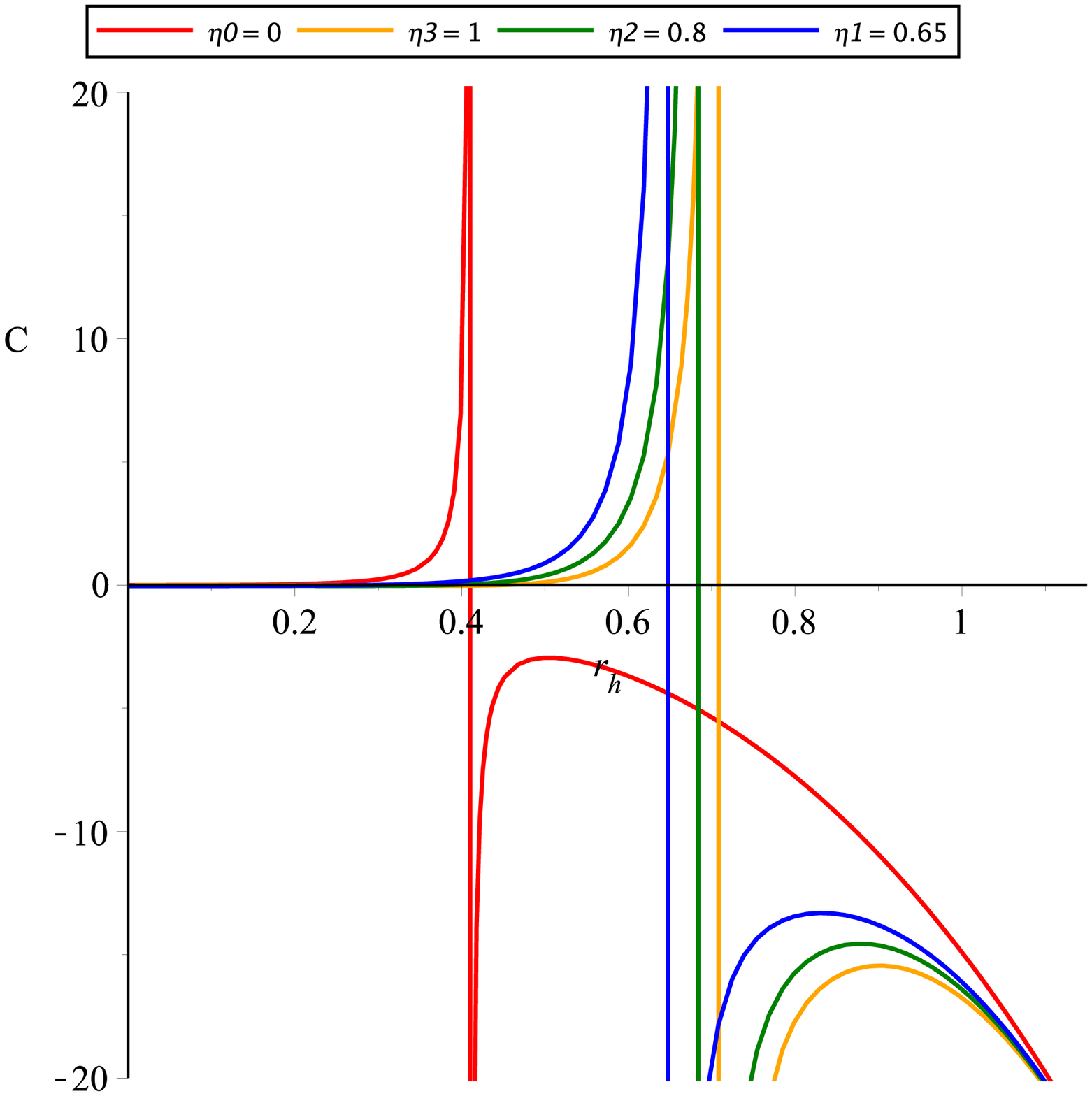}
	}
	\subfigure[closeup of figure (a) ]{
		\includegraphics[width=0.4\textwidth]{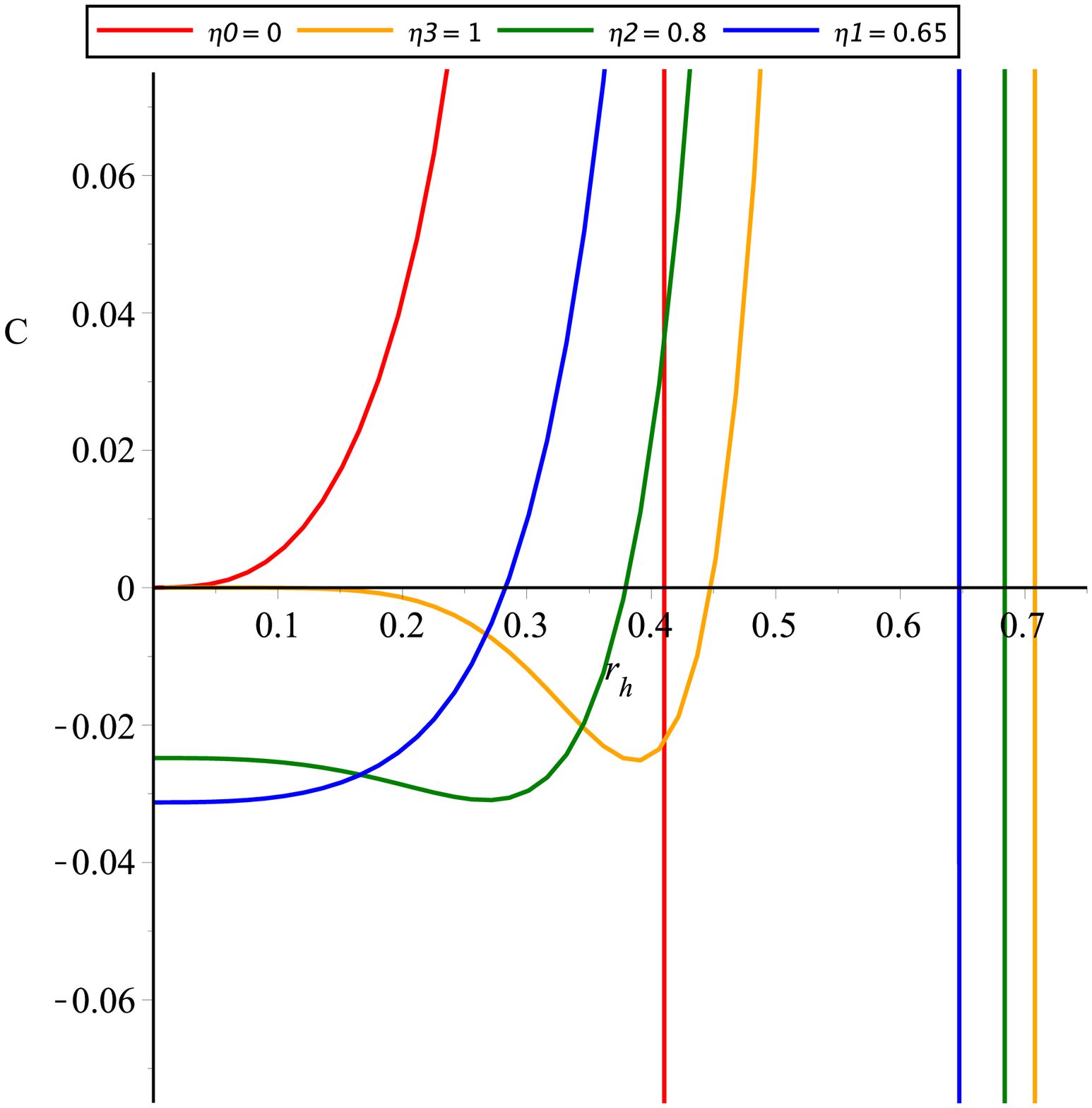}
	}
	\subfigure[$ \eta=0.95 $ ]{
		\includegraphics[width=0.4\textwidth]{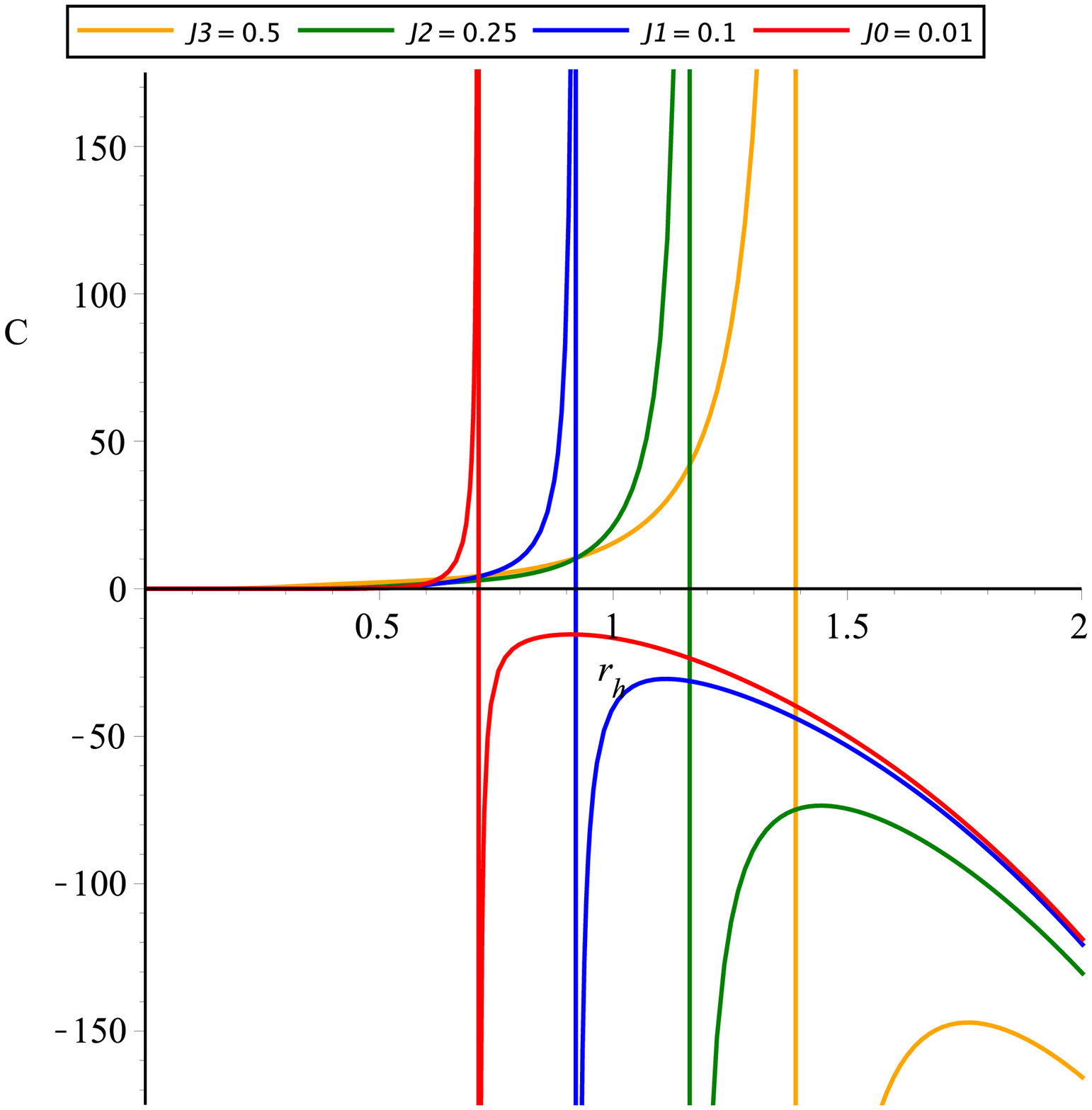}
	}
	\subfigure[closeup of figure (c) ]{
		\includegraphics[width=0.4\textwidth]{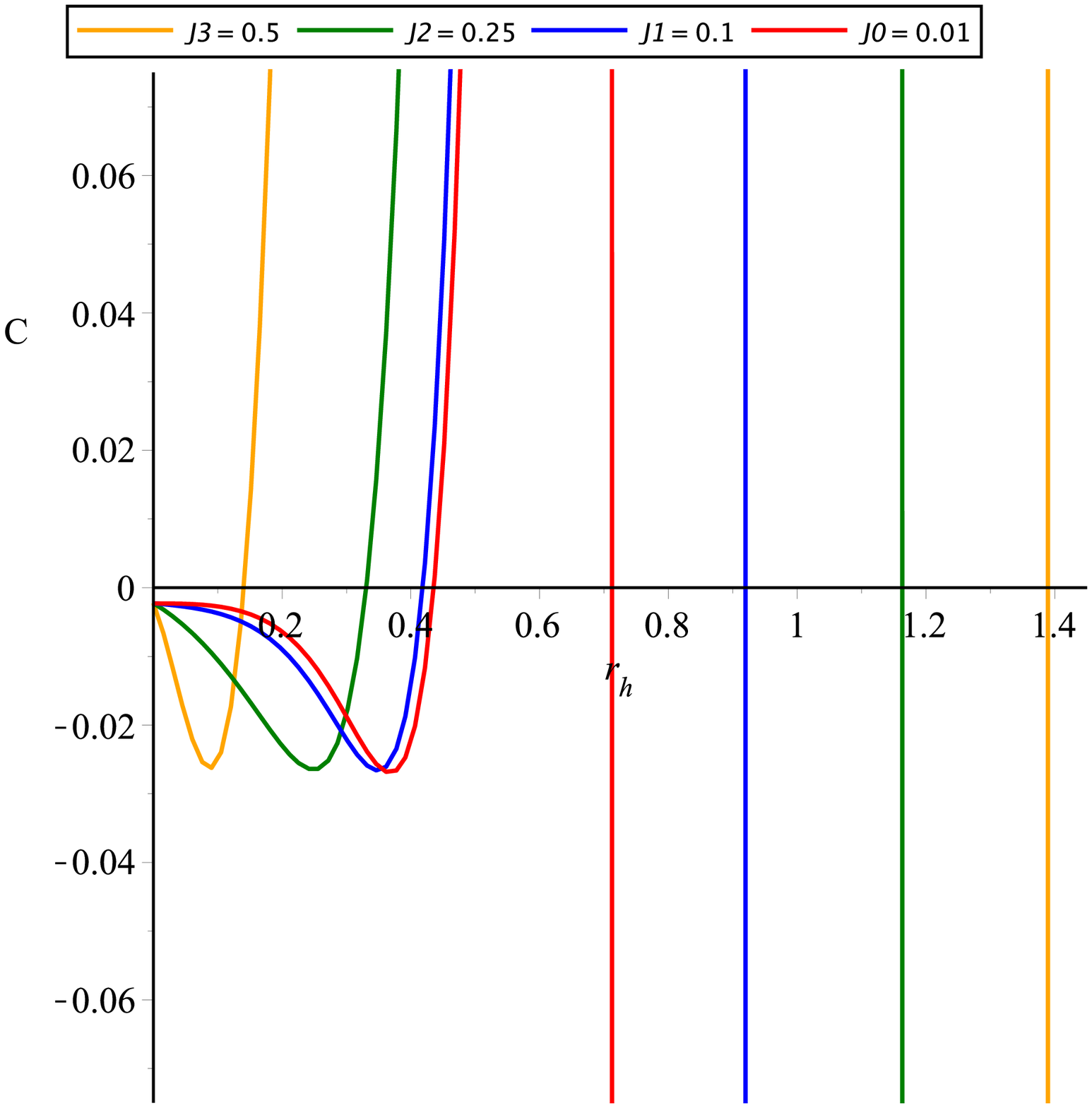}
	}
	\caption{Variations of the specific heat in terms of horizon radius $ r_{h} $, for the Myers-Perry black hole in five dimensions.}
	\label{pic:C}
\end{figure}

Such a quantum mass for a Myers-Perry black hole can be defined by first observing that the  original entropy  (without quantum gravitational corrections) of the   Myers-Perry black hole is  a function of original mass and angular momentum \cite{thermo, thermo1}. So, the original  mass can be expressed  as a function  of the original entropy and angular momentum. Now as the quantum gravitational corrections modify the original entropy, the quantum gravitationally corrected entropy can be used to define this novel quantum mass for the Myers-Perry black hole. Thus, the quantum mass can be expressed as  function of the quantum corrected entropy   
\begin{equation}\label{M-S-OMP}
M(S,J)=\frac{3}{8\pi^{4/3}}\left[16\pi^{4}J^{2}+4LW\left(-\frac{\eta}{e^{S}}\right)^{2}+8SLW\left(-\frac{\eta}{e^{S}}\right)+4S^2\right]^{\frac{1}{3}},
\end{equation}
where $LW\left(x\right)$ is Lambertw function.
Here, we can consider the  parameter $\eta$  as a new thermodynamic variable with conjugate $X$,  which is given by,
\begin{equation}\label{X}
X=\frac{\sqrt{96M^{3}-81J^{2}}}{32\pi^{2}M^{2}}e^{-\frac{2\pi^{2}\sqrt{96M^{3}-81J^{2}}}{9}}.
\end{equation}
Now,  we can obtain the non-perturbative corrections for other  thermodynamic quantities    as a function of $ S $  and $ J $. So,  we can write the quantum gravitationally corrected temperature   $(T=({\partial M}/{\partial S}))$ as
\begin{equation}_{\label{T}}
T={\frac {\xi_1}{8{\xi_2}^{2/3}{\pi }^{4/3}}}    ,
\end{equation}
We can also express the quantum gravitationally  corrected   specific heat $ (C= T({\partial S}/{\partial T})) $ as
\begin{equation}\label{C}
C={\frac {\xi_1}{8{\xi_2}^{2/3}{\pi }^{4/3}} \left( -{\frac {{
				\xi_1}^{2}}{12{g}^{5/3}{\pi }^{4/3}}}+{\frac {\xi_3}{8{g}^{2/3}{\pi }^
			{4/3}}} \right) ^{-1}},
\end{equation}
where
\begin{eqnarray}_{\label{T}}
\xi_1=-8\,{\frac { \left( LW \left( -\eta\,{{\rm e}^{-S}} \right)
		\right) ^{2}}{1+LW \left( -\eta\,{{\rm e}^{-S}} \right) }}+8\,LW
\left( \eta\,{{\rm e}^{-S}} \right) -8\,{\frac {S LW \left( \eta\,{
			{\rm e}^{-S}} \right) }{1+LW \left( \eta\,{{\rm e}^{-S}} \right) }}+8\,
S,
\\
\xi_2=16\,{\pi }^{4}{J}^{2}+4\, \left( LW \left( -\eta\,{{\rm e}^{-S}}
\right)  \right) ^{2}+8\,S LW \left( \eta\,{{\rm e}^{-S}} \right) +4\,{
	S}^{2},
\\
\xi_3=-16\,{\frac {LW \left( \eta\,{{\rm e}^{-S}} \right) }{1+LW \left(
		\eta\,{{\rm e}^{-S}} \right) }}+{\frac {8\,S LW \left( \eta\,{{\rm e}^{-
				S}} \right) +16\, \left( LW \left( -\eta\,{{\rm e}^{-S}} \right)
		\right) ^{2}}{ \left( 1+LW \left( -\eta\,{{\rm e}^{-S}} \right)
		\right) ^{2}}}  \nonumber\\
	-{\frac {8\, \left( LW \left( -\eta\,{{\rm e}^{-S}}
		\right)  \right) ^{3}+8\,S LW \left( \eta\,{{\rm e}^{-S}} \right) }{
		\left( 1+LW \left( -\eta\,{{\rm e}^{-S}} \right)  \right) ^{3}}}+8.
\end{eqnarray}
We have plotted the behavior of  these   thermodynamic quantities and investigated the effects of non-perturbative  gravitational quantum corrections on them.  Now from
 Fig.~\ref{pic:M}, Fig.~\ref{pic:T}, Fig.~\ref{pic:C}  and  we observe that  the quantum corrected mass, temperature and specific heat  depends on  $ \eta $ and $ J $. This is expected as $\eta$ is the parameter which measures the strength of non-perturbative quantum corrections. Also,  the angular momentum can change the effect of the non-perturbative quantum corrections on the thermodynamics of a Myers-Perry black hole. Here it has been demonstrated how the quantum gravitational corrections are first directly incorporated  in the quantum mass, and then their  effect on the angular momentum is obtain from this quantum mass.  This analysis can be generalized for charged black holes, and other  black hole solutions with  different topology. 
 As the information geometry of a black hole could be expressed using its mass,  we can use the quantum   mass to study the   quantum  gravitational  corrections to the information geometry for a quantum Myers-Perry black hole.   This will be done by  analyzing  the quantum gravitational corrections  to  different  information theoretical metrics    \cite{q1, q2, w1, w2, r1, r2, HPEM, h1, h2} of a quantum Myers-Perry black hole.  We   use different informational theoretical metrics to ensure that  we do not lose important  physical information  about the effects of the   non-perturbative quantum gravitational  corrections on phase transitions.  

We will start this analysis using the Ruppeiner metric.
The Ruppeiner metric was proposed to define a geometry for the space of   thermodynamical  fluctuations  \cite{r1, r2}. Now we can define the quantum corrected effective   Ruppeiner metric as
\begin{equation}
ds^2=   -\dfrac{1}{T}Mg_{ab}^{R}dX^{a}dX^{b} .
\end{equation}
The  Weinhold metric  was motivated using the connections between Gibbs-Duhem relation and scaling of thermodynamic potentials \cite{w1, w2}. So, we can write the  quantum corrected effective  Weinhold metric  as
\begin{equation}
ds^2=  Mg_{ab}^{W}dX^{a}dX^{b}.
\end{equation}
The  Quevedo (I and II) metrics   are constructed using Legendre invariant set of metrics in the phase space \cite{q1, q2} . It may be noted that their   pullback produces the required  information  metrics on the space of equilibrium states. Now motivated by the original  Quevedo (I and II)   metrics, we can define the quantum corrected effective   Quevedo (I and II) metrics  as
\begin{equation}
ds^2=(SM_{S}+\alpha M_{\alpha})(-M_{SS}dS^{2}+M_{\alpha\alpha}d\alpha^{2}),
\end{equation}
and
\begin{equation}
ds^2= \quad SM_{S}(-M_{SS}dS^{2}+M_{\alpha \alpha}d\alpha^{2}).
\end{equation}
The HPEM information metric  has been constructed by using  a  different conformal than the  Quevedo metrics \cite{HPEM, HPEM1, HPEM2, HPEM3}. Thus, the quantum corrected effective HPEM metrics  can  be  expressed as
\begin{equation}
ds^2=\quad \quad \dfrac{SM_{S}}{\left(\frac{\partial^{2} M}{\partial \alpha^{2}}\right)^{3}}\left(-M_{SS}dS^{2}+M_{\alpha\alpha}d\alpha^{2}\right).
\end{equation}
It is also possible to define NTG metric  by changing coordinates of the thermodynamic space by  Jacobean transformation  \cite{h1, h2}. Now  the quantum corrected effective NTG    metric can be written as
\begin{equation}
ds^2 = \quad \quad \dfrac{1}{T}\left(-M_{SS}dS^{2}+M_{\alpha\alpha}d\alpha^{2}\right).
\end{equation}
In these quantum corrected effective metrics,   $M_{XY}$ is the second order differentiation with respect to $X$ and $Y$ variables.  It may be noted that they have been obtained using the quantum corrected mass.\\

\begin{figure}[h]
	\centering
	\subfigure[]{
		\includegraphics[width=0.45\textwidth]{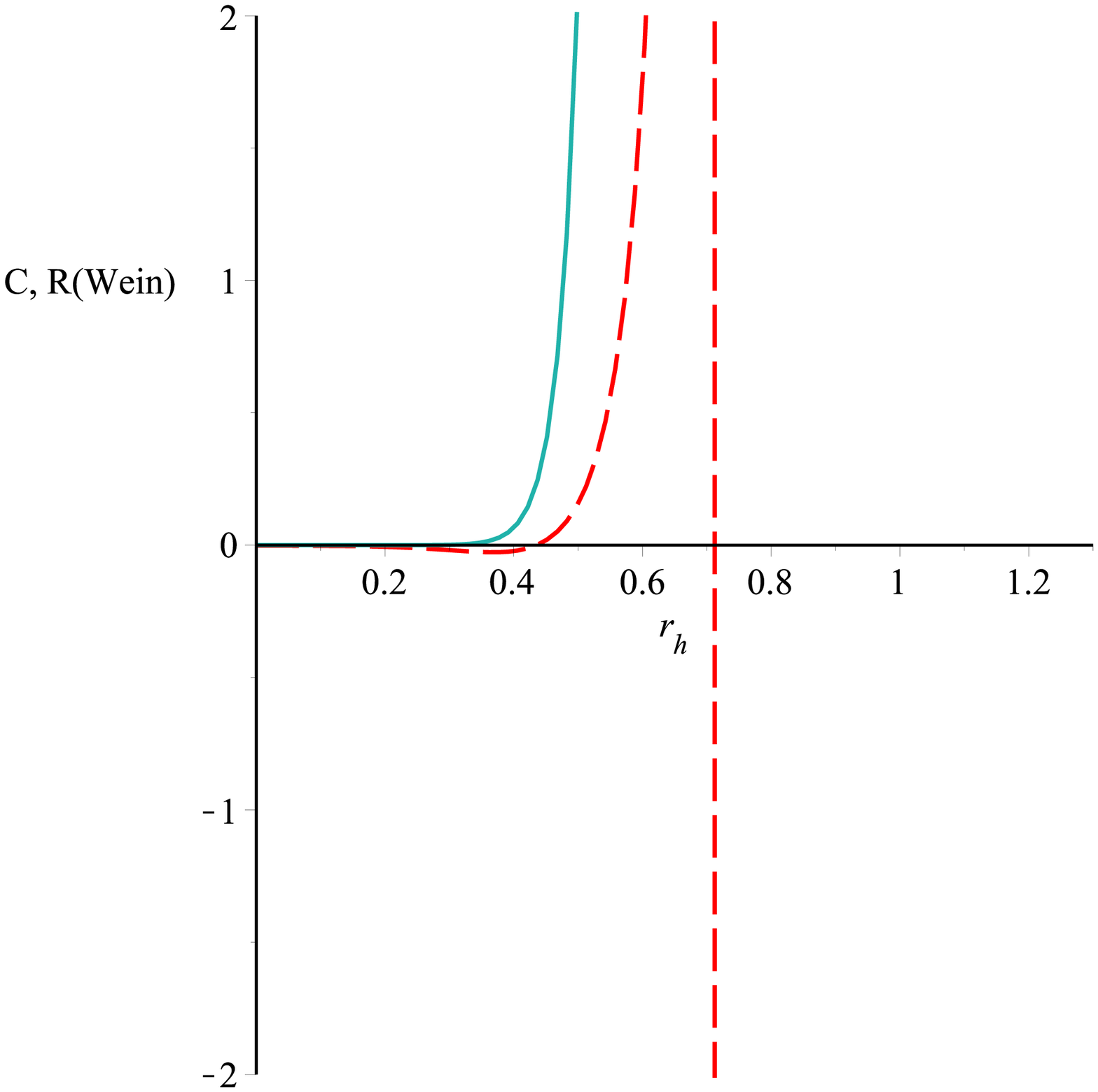}
	}
	\subfigure[]{
		\includegraphics[width=0.45\textwidth]{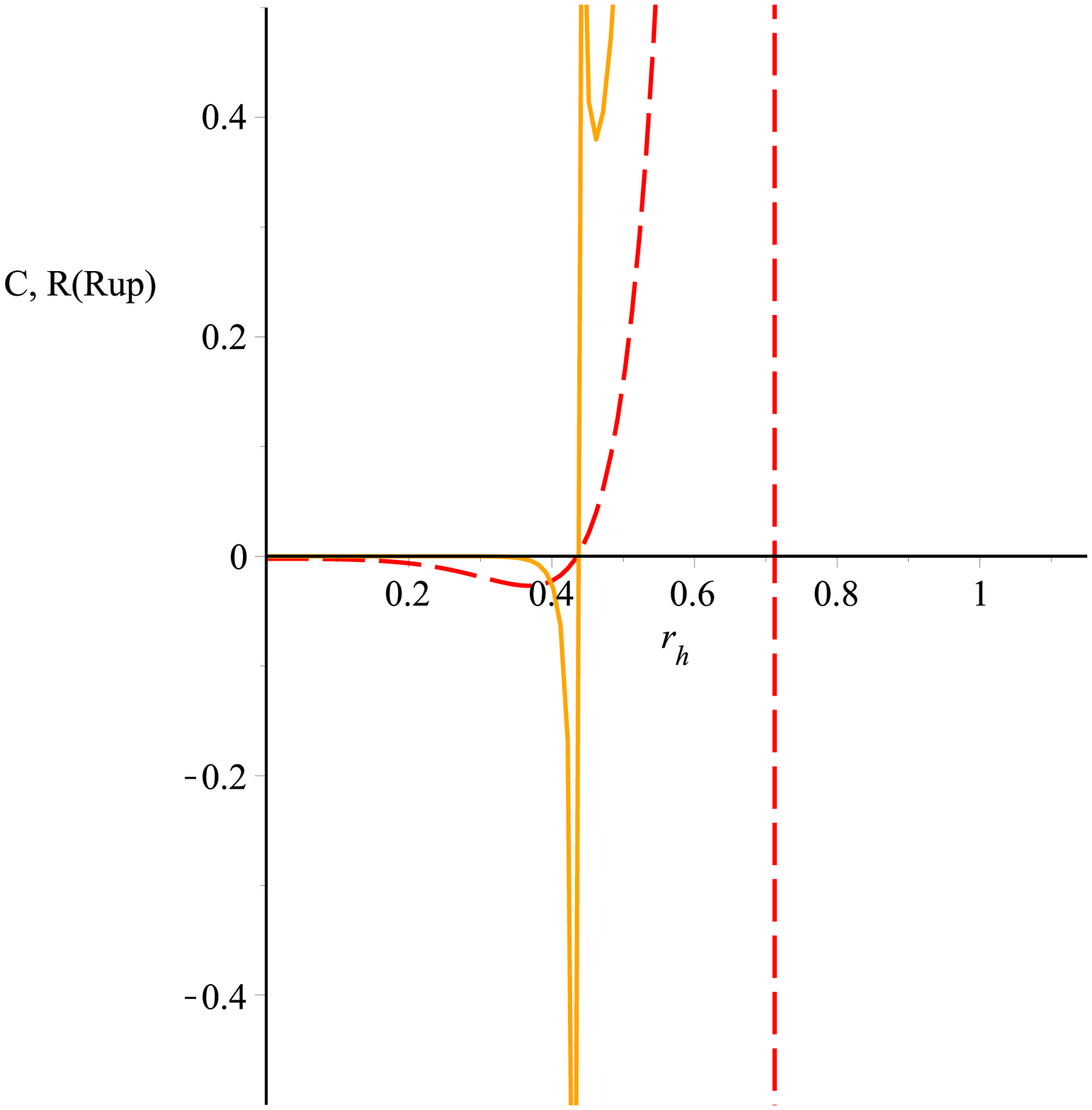}
	}
	\subfigure[]{
		\includegraphics[width=0.45\textwidth]{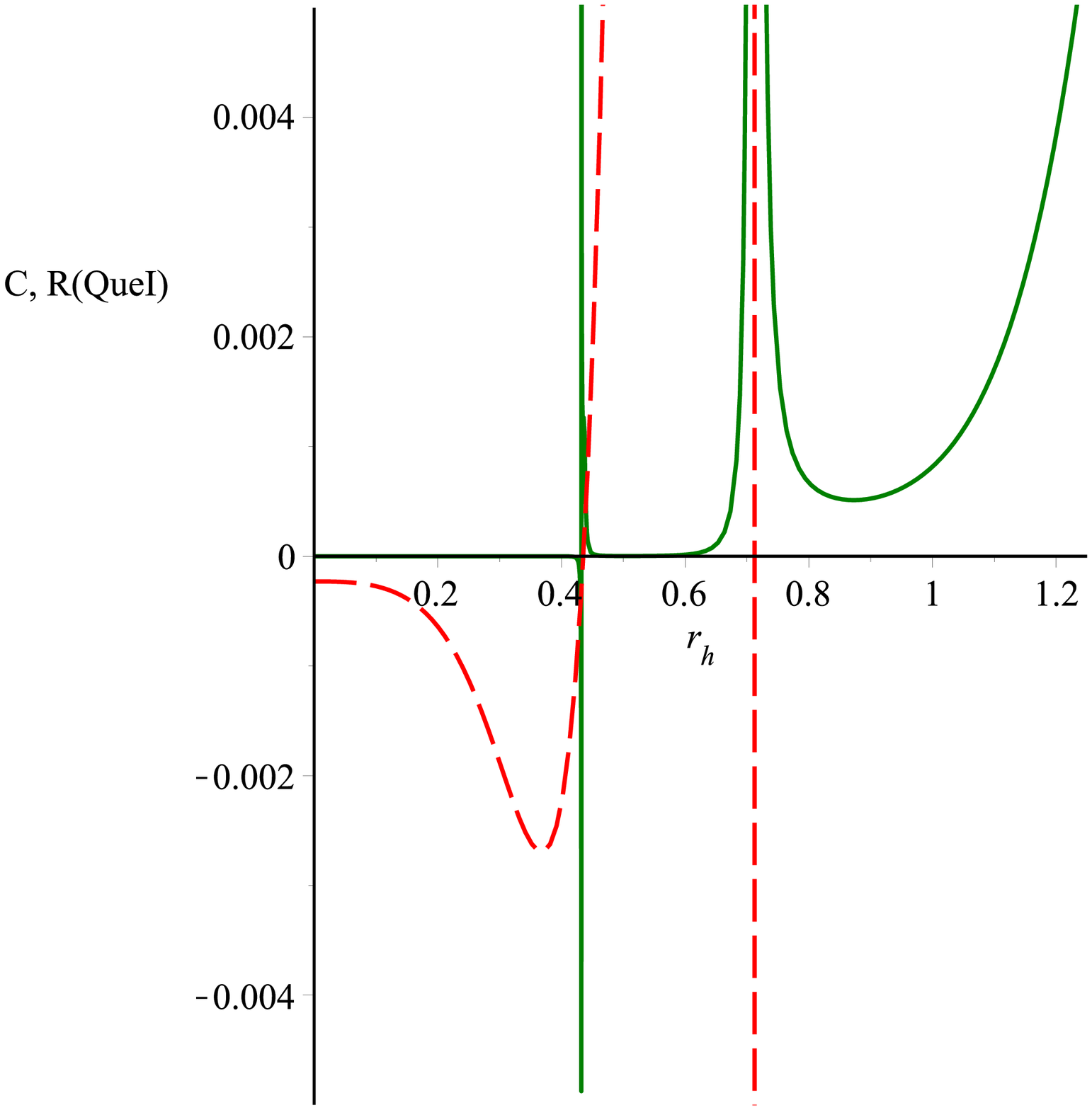}
	}
	\subfigure[]{
		\includegraphics[width=0.5\textwidth]{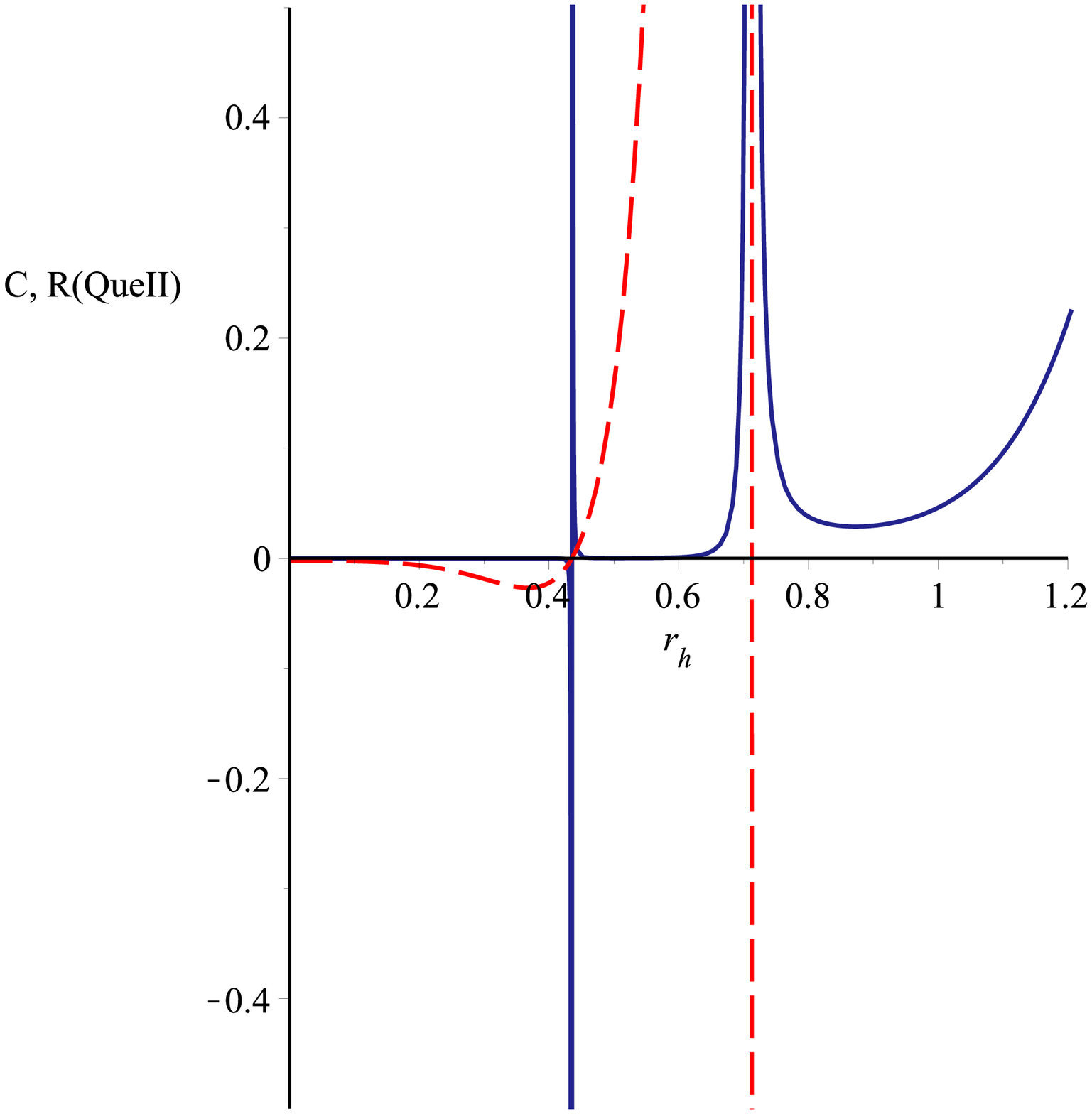}
    }	
	\caption{Curvature scalar variation of Weinhold (light green line), Ruppeiner (orange line), Quevedo case I (green line) and Quevedo case II (navy line) metrics and also specific heat  (red dash line), in terms of horizon radius $ r_{h} $ for $ \eta=0.95 $ and $J=0.01 $, for a Myers-Perry black hole. }
	\label{pic:CRWeinQue}
\end{figure}

\begin{figure}[h]
	\centering

	\subfigure[]{
		\includegraphics[width=0.45\textwidth]{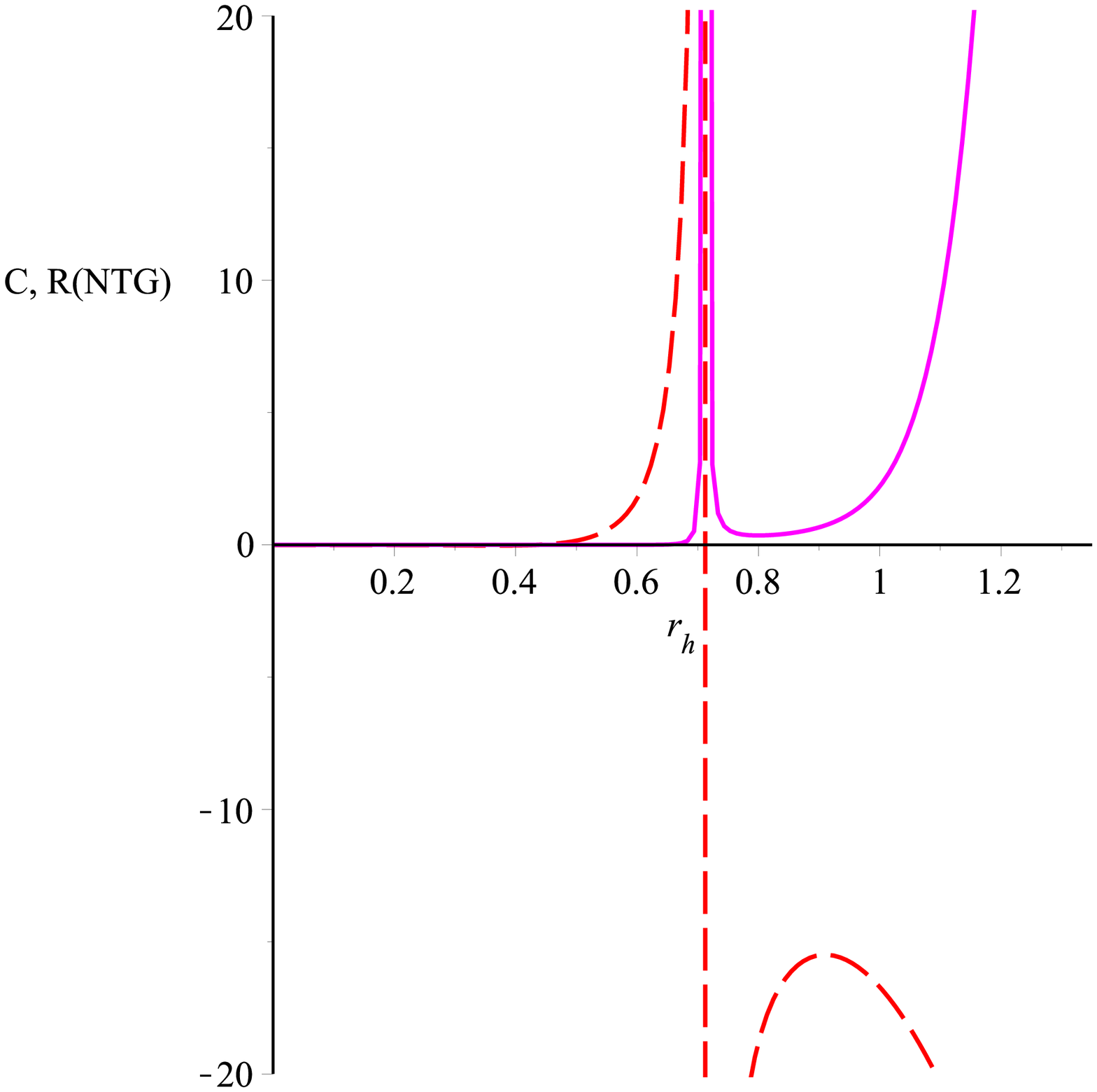}
	}
	\subfigure[closeup of figure (a)]{
		\includegraphics[width=0.45\textwidth]{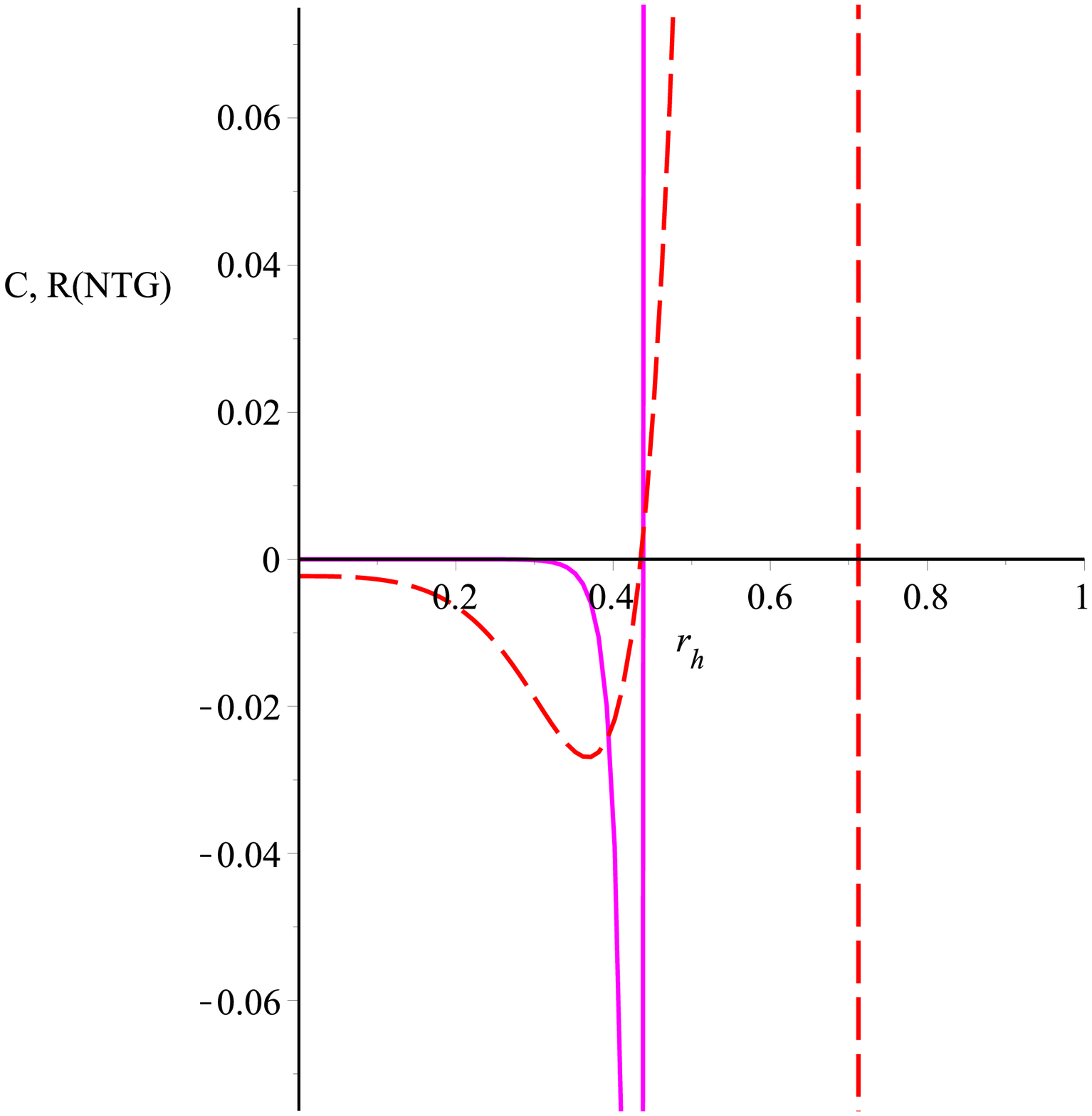}
	}
    \subfigure[]{
    	\includegraphics[width=0.45\textwidth]{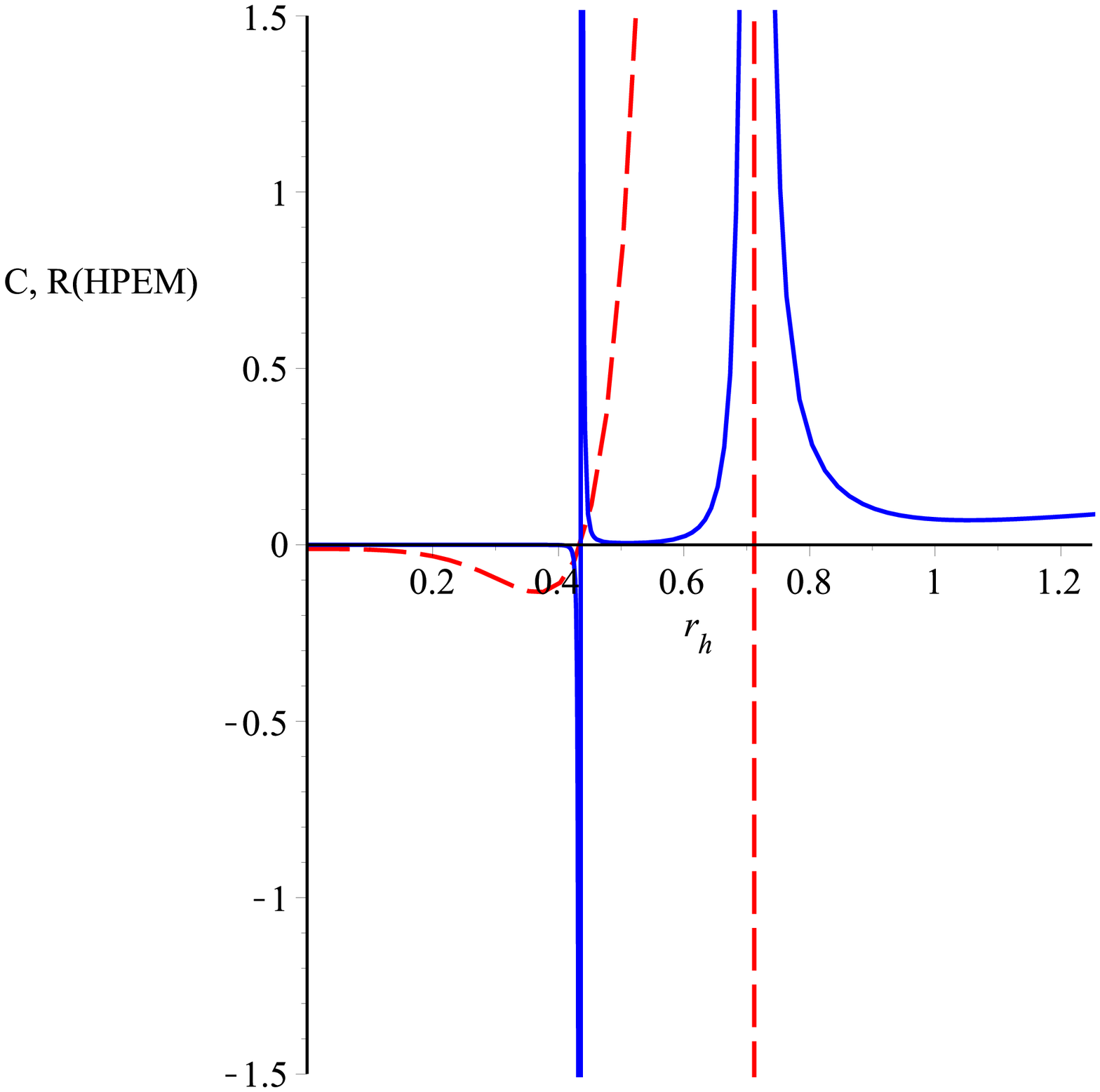}
    }
	\caption{ Scalar curvature  of NTG (magenta line) and HPEM (blue line) metrics  and also specific heat  (red dash line), in terms of horizon radius $ r_{h} $ for $ \eta=0.95 $ and $J=0.01 $, for  a Myers-Perry black hole.}
	\label{pic:CRHPEMNTG}
\end{figure}

Now, we can use plots to investigate the behavior of quantum corrected effective Ricci scalars,  and compare it with  the behavior of the specific heat (see Figs.~\ref{pic:CRWeinQue}, \ref{pic:CRHPEMNTG}).
We find that quantum corrected effective Weinhold metric does not have any   divergences,  where the specific heat  diverges or becomes  zero (see Figs.~\ref{pic:CRWeinQue} (a)). Furthermore,   from Fig.~\ref{pic:CRWeinQue} (b), we observe that  the quantum corrected effective Ruppeiner metric has a divergence,  where the specific heat becomes zero. So, the  divergence of the scalar curvature  for the quantum corrected effective Ruppeiner metric  corresponds to a phase transition in this system.
The situation is different for  the  quantum corrected effective  Quevedo  (I, II), quantum corrected effective HPEM, and quantum corrected effective NTG metrics. The divergences of the  scalar curvature  of  these quantum corrected effective  metrics can provide better information about the  phase transition points for a Myers-Perry black hole (see Figs.~\ref{pic:CRWeinQue} (c), (d) and \ref{pic:CRHPEMNTG} (a)--(d)).
In fact, we observe that the divergent points of the quantum corrected effective Ricci scalar for the quantum corrected effective Quevedo ( I, II), quantum corrected effective HPEM and quantum corrected effective  NTG metrics contain important information about   phase transitions type in this system. The divergent points of the quantum corrected effective Ricci scalar for these quantum corrected effective  metrics coincide with both zero  and divergent points of the  specific heat. So,  these quantum corrected effective metrics contain more information about the phase transition in this system.  Thus,    quantum corrected effective Quevedo  (I, II), quantum corrected effective HPEM and quantum corrected effective NTG metrics can be used to obtain important physical information about the phase transition in a quantum Myers-Perry black hole. Here we observe that different quantum corrected informational metrics also contain different amounts of information about the system.  It  is interesting to note that the information geometries have been constructed by incorporating quantum gravitational corrections  in the effective informational metrics, which were obtained from a novel quantum mass of a Myers-Perry black hole. 
\section{Conclusion}
In this paper, we analyzed quantum work for a quantum scale  Myers-Perry black hole. It was observed that the quantum work was corrected by  non-perturbative correction for a quantum scale  Myers-Perry black hole.  As these  corrections changed the relation between entropy and area of a Myers-Perry black hole,  they also  produces corrections in free energy of a Myers-Perry black hole.
So,  the difference of corrected free energy was used to obtain the corrections to the  quantum work.  We  used the  Jarzynski equality  to obtain the corrected   quantum work  for this black hole.

It may be noted that it is possible to use the original entropy $S_0$ of the black hole to obtain quantum work. As the original entropy $S_0 = A/4$ is obtained by semi-classical approximations (using quantum field theory in curved space-time), the equilibrium thermodynamic quantities obtained from $S_0$ correspond to this approximation \cite{1a, 2a}. Now as the difference between such  equilibrium free energies  can be related to quantum work using the  Jarzynski equality, the  Jarzynski equality has been used to obtain the work corresponding to the original entropy $S_0$ of black holes \cite{rz12, rz14}. However, for quantum scale black holes, it has been argued that non-perturbative quantum gravitational  corrections will correct this original entropy   \cite{2007.15401, Dabholkar, ds12, ds14}.  These non-perturbative quantum  gravitational corrections produce corrections to the original semi-classical entropy $S_0$, which can be represented as  $S = S_0 + \eta \exp (-S_0) $  \cite{2007.15401, Dabholkar, ds12, ds14}.  Now  it is possible to  obtain the quantum gravitational corrections to the  free energies from this quantum corrected entropy, we obtain quantum gravitational corrections  to the quantum work for a Myers-Perry black hole.    These quantum gravitational  corrections are controlled by a parameter, and the behavior of the quantum work explicitly depended on this control parameter. It is possible to use the mircostates of a  black hole to construct its partition function. As the black hole evaporates, its microstates  change, and this changes the partition function of the system. We obtained the relative weights of such partitions functions using the Jarzynski equality for a quantum   Myers-Perry black hole. 
Unlike heat represented by Hawking radiation, this   quantum work is represented by a  unitary information preserving process. So, there is an unitary  information preserving process associated with the evaporation of black holes, and this can have implication for black hole information paradox. 
As quantum work was done during the evaporation of a Myers-Perry black hole, we also investigated the effect of such corrections on the stability of a Myers-Perry black hole. It was observed that even though the original  Myers-Perry black hole become stable at short distance, the non-perturbative quantum corrections changed this behavior.  These corrections depended on the angular momentum of the black hole. So, in absence of angular momentum, these corrections  stabilized  the quantum scale Myers-Perry   black hole.
We analyzed the quantum  corrected information geometry of these black holes with different quantum corrected effective information metrics. It was observed that quantum corrected effective  Quevedo  (I, II), quantum corrected effective HPEM and quantum corrected effective NTG metrics  contain more  information about the phase transition in this system, than the quantum corrected effective Weinhold and quantum corrected effective  Ruppeiner metrics. These results   obtained by incorporating quantum gravitational corrections  in  effective informational metrics, which was obtained using a novel quantum mass of a Myers-Perry black hole.

It is known that the entropy of a back hole can be obtained from microstates  of a conformal field theory. The microstates  of black holes with angular momentum  have also been studied using the AdS/CFT correspondence. Thus, it is possible to analyze the microstates  of a rotating black hole using a conformal field theory dual to such a system.
Furthermore, in this paper, we could derive the expression for work done during the evaporation of a Myers-Perry black hole. This was done using the Jarzynski equality  for such a system. It is possible to investigate this using the AdS/CFT correspondence, and find the field theory dual to such work done during the evaporation of a black hole.  We could  analyze the quantum corrected information geometry of black holes using the AdS/CFT correspondence. It is expected that these different information theoretical metrics would be produce interesting dual structures for a Myers-Perry black hole. It is possible  to extend this work to other black holes. These are solutions  for an  extended objects surrounded by event horizons. It is known that these exotic solutions still have the   unusual causal structure.  It is also possible to relate the black strings to fundamental strings of the string theory. It has been known that these black strings can have the usual properties associated with black holes. Thus, it is possible to analyze the thermodynamics of black strings, and even use the information theoretical metrics to analyze phase transitions in black strings. So, it would be interesting to investigate the effect of these non-perturbative quantum corrections on the thermodynamics of black strings. It is possible   to investigate this modified thermodynamics of black strings using different information theoretical metrics. We can also obtain the quantum work for black stings using the Jarzynski equality for such a system. We could  explicitly obtain the quantum work for black strings using the  difference of their free energies.

It is known that in the Jacobson formalism, space-time emerges from the thermodynamics of the system. Now, it has been observed that the thermodynamics is corrected due to non-perturbative quantum corrections. So, it is possible to use this quantum corrected thermodynamics for the Myers-Perry   black hole to obtain a quantum corrected effective metric for such black holes. These quantum corrections in this effective metric are expected  be proportional to the coefficient of the non-perturbative corrections. Thus, at large distances, this quantum corrected geometry of a  Myers-Perry   black hole could be approximated by its classical metric. However, the quantum corrections to the metric would become important at short distances. It would be interesting to investigate the Raychaudhuri equation for such
quantum gravitationally corrected effective geometries. This could then be used to investigate the effect of quantum corrections on the singularity theorems. It is expected that the geometry flow would also be modified by such non-perturbative quantum corrections. So, it   would also be interesting to investigate what such effects for a  Myers-Perry  black hole, when  its thermodynamics is modified from non-perturbative quantum  corrections.  

As we analyzed the effect of quantum gravitational corrections on a  Myers-Perry  black hole, it would be also interesting to analyze quantum work for fuzzballs.
It is possible to analyze the black holes in string theory with fuzzball proposal. In this proposal, the singularity at the center of a black hole is smoothed out. This is done by assuming that the entire region inside the horizon of a  black hole
is made of string states, and these string states resemble a black hole. As the effective classical description of a fuzzball still resembles a black hole, it is consistent with large scale  gravitational experiments conducted on large scale astrophysical black holes. The large scale thermodynamics from fuzzballs also resembles classical black holes.   However, we do expect that short distances physics of fuzzballs to be different from black holes. So, it would be interesting to investigate the effect of quantum corrections on fuzzballs. It would also be possible to obtain the quantum work for fuzzballs by analyzing quantum thermodynamics of fuzzballs. It is also expected that the quantum work of fuzzballs would be corrected from short distance quantum corrections. It is possible  to investigate such corrections to quantum work for fuzzballs using the Jarzynski equality. Thus, we can first try to obtain the equilibrium thermodynamic quantities for a fuzzball. The difference of two equilibrium free energies at two different states can then be related to the average  quantum work. In fact,   this average quantum work is related to the different weights of the partition functions, and  the partition functions can be calculated in a fuzzballs in terms of string microstates. So, it would be interesting to related such partition functions using quantum work for a fuzzball.


\begin{thebibliography}{99}

\bibitem{1a}S. W. Hawking, Nature 248 (1974) 30.
\bibitem{2a}S. W. Hawking, Commun. Math. Phys. 43  (1975) 199.
\bibitem{4a}L. Susskind, J. Math. Phys. 36  (1995)  6377.
\bibitem{5a} R. Bousso, Rev. Mod. Phys. 74  (2002) 825.
\bibitem{6ab}G. Lifschytz and M. E. Ortiz, Nucl.Phys. B 486 (1997) 131. 

\bibitem{7ab}S.~Mahapatra, Eur. Phys. J. C  {78}    (2018) 23. 
\bibitem{6ba}K.~Nozari and A.~S.~Sefiedgar,
Gen. Rel. Grav.  501 (2007) 39. 
\bibitem{7ba}C.~Keeler, F.~Larsen and P.~Lisbao,
Phys. Rev. D   043011 (2014) 90. 

\bibitem{6a}D. Bak and S. J. Rey, Class. Quant. Grav. 17  (2000) L1.
\bibitem{7a} S. K. Rama, Phys. Lett. B 457  (1999) 268.


\bibitem{18}
S. Hemming and L. Thorlacius,   JHEP 11 (2007) 086.
\bibitem{18a}R.~Gregory, S.~F.~Ross and R.~Zegers,
 JHEP  {09}  (2008) 029.
\bibitem{18b} J.~V.~Rocha, JHEP {08}  (2008) 075.
\bibitem{18c}Z.~H.~Li, B.~Hu and R.~G.~Cai, Phys. Rev. D {77} (2008)
104032.
\bibitem{18d} K.~Saraswat and N.~Afshordi,
 JHEP  136 (2020) {04}.

\bibitem{19}
R. B. Mann and S. N. Solodukhin,   Nucl. Phys. B 523 (1998) 293.
\bibitem{19a}A.~Sen, Gen. Rel. Grav. {44} (2012) 1947.


\bibitem{Ashtekar}
A. Ashtekar, \textit{Lectures on non-perturbative canonical gravity}, World Scientific: Singapore (1991).


\bibitem{Govindarajan}
T. R. Govindarajan, R. K. Kaul and V. Suneeta,  Class. Quantum Grav. 18 (2001) 2877.
\bibitem{29}
D. Birmingham and S. Sen, Phys. Rev. D 63 (2001) 047501.
\bibitem{gr12}T. Jacobson, Phys. Rev. Lett. 75 (1995) 1260.
\bibitem{gr14} M.~Faizal, A.~Ashour, M.~Alcheikh, L.~Alasfar, S.~Alsaleh and A.~Mahroussah, Eur. Phys. J. C {77} (2017)  608.

\bibitem{32}
S. Das, P. Majumdar and R. K. Bhaduri,   Class. Quantum Grav. 19 (2002) 2355.
\bibitem{32a}S. Upadhyay, B. Pourhassan and H. Farahani,
  Phys. Rev. D 95 (2017) 106014.
\bibitem{32b} A.~Jawad, Class. Quant. Grav.  {37} (2020)   185020.
\bibitem{32c}B.~Pourhassan, Eur. Phys. J. C {79} (2019)   740.
\bibitem{32d} J.~Sadeghi, B.~Pourhassan and M.~Rostami,
Phys. Rev. D  {94} (2016)  064006.
\bibitem{40a}B.~Pourhassan, S.~Dey, S.~Chougule and M.~Faizal,
 Class. Quant. Grav.  {37} (2020) 135004.
\bibitem{40b}B.~Pourhassan, S.~Upadhyay, H.~Saadat and H.~Farahani,
 Nucl. Phys. B {928} (2018) 415.
\bibitem{40c}B.~Pourhassan, M.~Faizal, Z.~Zaz and A.~Bhat,
 Phys. Lett. B  {773} (2017) 325.
\bibitem{40d} B.~Pourhassan, A.~Ovgun and \.I.~Sakalli,
 Int. J. Geom. Meth. Mod. Phys.  {17} (2020)  2050156




\bibitem{2007.15401}
A. Chatterjee and A. Ghosh,   Phys. Rev. Lett. 125 (2020) 041302.
\bibitem{Dabholkar}
A. Dabholkar, J. Gomes and S. Murthy,   JHEP 03 (2015) 074.
\bibitem{ds12}A.~Dabholkar, J.~Gomes and S.~Murthy, JHEP 04  (2013) 062. 
\bibitem{ds14} S. Murthy and B. Pioline,  JHEP 09 (2009) 022. 
\bibitem{MP}
R. C. Myers and M. J. Perry,   Annals Phys. 172  (1986)
304. 

\bibitem{thermo}D.~Astefanesei, M.~J.~Rodriguez and S.~Theisen, JHEP {08}, 046 (2010)
\bibitem{thermo1} M.~Garbiso and M.~Kaminski, JHEP {12}, 112 (2020)

\bibitem{thermo2}H.~Saadat and A.~Pourdarvish,
Int. J. Theor. Phys.  {53},   (2014) 3014. 
\bibitem{thermo4} M.~Stein, JHEP {09}, 067 (2016)
\bibitem{thermo5}B.~P.~Dolan,
Phys. Rev. D  {92}   (2015) 044013. 

\bibitem{temp}D.~F.~Litim and K.~Nikolakopoulos,
 JHEP  {04} (2014), 021
 
 \bibitem{10th}B. de L. Bernardo, Phys. Rev. E 102 (2020) 062152. 
 
 \bibitem{12th}A.~Teixid\'o-Bonfill, A.~Ortega and E.~Mart\'\i{}n-Mart\'\i{}nez, Phys. Rev. A  {102} (2020)  052219.
 \bibitem{12tha}A.~Ortega, E.~McKay, \'A.~M.~Alhambra and E.~Mart\'\i{}n-Mart\'\i{}nez, Phys. Rev. Lett. {122} (2019)   240604. 
 \bibitem{paradox1}S.~L.~Braunstein and A.~K.~Pati,
Phys. Rev. Lett. {98} (2007), 080502. 
 
 \bibitem{paradox2}D. N. Page, Phys. Rev. Lett. 71 (1993) 1291.  

\bibitem{info1}J.~Liu, H.~Yuan, X.~M.~Lu and X.~Wang,
 J. Phys. A {53} (2020)  023001.
\bibitem{info2}J. Goold, M. Huber, A. Riera, L. del Rio and P. Skrzypczyk, 	J. Phys. A  49  (2016) 143001.


\bibitem{paradox4}S.~Lloyd and J.~Preskill, JHEP {08} (2014) 126. 
\bibitem{paradox5}D. Marolf and J. Polchinski
Phys. Rev. Lett. 111 (2013) 171301. 

\bibitem{mi12}S. Carlip, Class. Quant. Grav. 17 (2000) 4175.
\bibitem{mi14}J.~l.~Jing and M.~L.~Yan,
Phys. Rev. D 63 (2001) 024003.
\bibitem{mi16}M.~Hassaine,
Phys. Rev. D  101   (2020) 084028. 
\bibitem{mi18}M.~Honda, Phys. Rev. D  {100}   (2019) 026008.   
\bibitem{ro12} B. Chen and J. Long,  JHEP 1006 (2010) 018. 
\bibitem{ro14}S.~M.~Noorbakhsh and M.~Ghominejad,
Phys. Rev. D  {95}   (2017) 046002. 
\bibitem{ro16}A.~Pourdarvish and B.~Pourhassan,
Int. J. Theor. Phys.  {53} (2014) 136. 
\bibitem{ro18} 
B.~Chen, S.~X.~Liu and J.~J.~Zhang, JHEP {11} (2012) 017. 
%%%%%%%%%%%%%%%%%%%%%%%%%%%%%%%%%%%%
\bibitem{adba0}Z. Fei, N. Freitas, V. Cavina, H. T. Quan and  M.  Esposito,  Phys. Rev. Lett. 124 (2020)  170603. 
\bibitem{adba1} B.~B.~Wei, Phys. Rev. E  {97}   (2018) 012114. 
\bibitem{adba2}J.~Salmilehto, P.~Solinas and M.~Mottonen, 
Phys. Rev. E  {89}   (2014) 052128.
\bibitem{adba4}P. Talkner, E.  Lutz and P. Hanggi,  Phys. Rev. E 75 (2007) 050102.
 
\bibitem{work1}Crooks, J. Stat. Phys. 90 (1998) 1481.  
\bibitem{work2} G. E. Crooks, Phys. Rev.
E 60 (1999) 2721. 
\bibitem{eq12}C. Jarzynski,  Phys. Rev. Lett.
78 (1997) 2690. 
\bibitem{eq14} C. Jarzynski,  J. Stat. Phys. 96 (1999) 415. 

\bibitem{rz12}S.~Iso, S.~Okazawa and S.~Zhang,
Phys. Lett. B  {705}  (2011) 152. 
\bibitem{rz14}S.~Iso and S.~Okazawa,
Nucl. Phys. B {851}  (2011) 380. 


\bibitem{1111}H. Dimov, R. C. Rashkov, T. Vetsov, Phys. Rev. D 99 (2019) 126007.

\bibitem{2222}T. Vetsov, Eur. Phys. J. C 79 (2019) 71.

\bibitem{point1}A.~Sheykhi, F.~Naeimipour and S.~M.~Zebarjad, Phys. Rev. D 91  (2015) 124057. 
\bibitem{point2} G.~Q.~Li and J.~X.~Mo, Phys. Rev. D 93   (2016) 124021. 
\bibitem{point4}S-W. Wei  and Y-X. Liu 
Phys. Rev. Lett. 115  (2015) 111302. 
\bibitem{point5} S-W. Wei, Y-X. Liu and R. B. Mann
Phys. Rev. Lett. 123 (2019) 071103.
\bibitem{badpa}M. Dehghani and M. Badpa, Prog. Theor. Exp. Phys. 2020(17) (2020) 033E03.
\bibitem{3dmpl}M. Dehghani, Phys. Lett. B, 803 (2020) 135335.
 \bibitem{12qb} S.~Soroushfar and S.~Upadhyay, Phys. Lett. B {804} (2020) 135360. 
\bibitem{12ra}A. Ghosh and  C.  Bhamidipati, Phys. Rev. D 101 (2020) 046005. 
\bibitem{12rw}J.~Suresh, R.~Tharanath, N.~Varghese and V.~C.~Kuriakose, Eur. Phys. J. C {74} (2014) 2819. 

 
\bibitem{rh12} S.~Soroushfar, R.~Saffari and S.~Upadhyay, Gen. Rel. Grav. 51  (2019) 130. 



\bibitem{r1} G. Ruppeiner, Rev. Mod. Phys. 67 (1995) 605.
\bibitem{r2} G. Ruppeiner, Phys. Rev. A 20 (1979) 1608.


\bibitem{w1}  F. Weinhold, J. Chem. Phys. 63 (1975) 2484.
\bibitem{w2}  F. Weinhold, J. Chem. Phys. 63 (1975) 2479.

\bibitem{q1}  H. Quevedo, JHEP  09 (2008) 034.
\bibitem{q2} H. Quevedo, J. Math. Phys. 48 (2007) 013506.

\bibitem{HPEM} S. H. Hendi, B. Eslam Panah, S. Panahiyan and M. Momennia, Eur. Phys. J. C 75 (2015) 507.

\bibitem{HPEM1}S. H. Hendi, B. Eslam Panah, and S. Panahiyan, JHEP  05 (2016) 029.
\bibitem{HPEM2} S. H. Hendi, A. Sheykhi, S. Panahiyan and B. Eslam Panah, Phys. Rev. D 92 (2015) 064028.
\bibitem{HPEM3} S. H. Hendi, S. Panahiyan, B. Eslam Panah and Z. Armanfard, Eur. Phys. J. C, 76 (2016) 396.
 
 
\bibitem{h1}
S. A. H. Mansoori and B. Mirza, Phys. Lett. B 799  (2019) 135040.
\bibitem{h2}
S.~A.~H. Mansoori, M.~Rafiee and S.~W.~Wei,
Phys. Rev. D  {102}   (2020) 124066. 


\end{thebibliography}
\end{document}